\documentclass[12pt,preprint,letterpaper,xdvi]{aastex}
\usepackage{amsmath} 
\usepackage{wrapfig} 
\usepackage{array}  
\usepackage{natbib}
\usepackage{brief_bib}
\usepackage{lscape}
\usepackage{verbatim}
		
\newcommand\bvec{{\bf B}}
\newcommand\avec{{\bf A}}

\newcommand\nhat{{\bf \hat{n}}} 

\newcommand\rvec{{\bf r}}

\newcommand\vvec{{\bf v}}

\newcommand\al{$\alpha$}
\newcommand\als{$\alpha\mbox{ }$}

\begin{document}

\title{Direct Measurements of Magnetic Twist in The Solar Corona}
\author{A.~Malanushenko}
\affil{Department of Physics, Montana State University\\
Bozeman, MT 59717, USA}
\author{M.~H.~Yusuf} 
\affil{Berea College, Berea, KY 40404, USA}
\author{D.W.~Longcope}
\affil{Department of Physics, Montana State University\\
Bozeman, MT 59717, USA}

\begin{abstract}
In the present work we study evolution of magnetic helicity in the solar corona. We compare the rate of change of a quantity related to the magnetic helicity in the corona to the flux of magnetic helicity through the photosphere and find that the two rates are similar. This gives observational evidence that helicity flux across the photosphere is indeed what drives helicity changes in solar corona during emergence. 

For the purposes of estimating coronal helicity we neither assume a strictly linear force-free field, nor attempt to construct a non-linear force-free field. For each coronal loop evident in Extreme Ultraviolet (EUV) we find a best-matching line of a linear force-free field and allow the twist parameter \als to be different for each line. This method was introduced and its applicability was discussed in \citet{Malanushenko2009}. 

The object of the study is emerging and rapidly rotating AR 9004 over about 80 hours. As a proxy for coronal helicity we use the quantity $\langle\alpha_iL_i/2\rangle$ averaged over many reconstructed lines of magnetic field. We argue that it is approximately proportional to ``flux-normalized'' helicity $H/\Phi^2$, where $H$ is helicity and $\Phi$ is total enclosed magnetic flux of the active region. The time rate of change of such quantity in the corona is found to be about $0.021$ rad/hr, which is compatible with the estimates for the same region obtained using other methods \citep{Longcope2007b}, who estimated the flux of normalized helicity of about $0.016$ rad/hr.
\end{abstract}

\section{Introduction}
\label{sec_intro}

Magnetic helicity is generally accepted to be an important quantity in understanding evolution of coronal magnetic fields and studying solar eruptions. It is approximately conserved when conductivity is high, as it is expected to be in solar corona. This sets up an important constraint on evolution of a magnetic field. It is believed that helicity in excess of a certain threshold could be responsible for triggering magnetohydrodynamics instabilities and thus be responsible for coronal mass ejections (CMEs) \citep[see][for further discussion]{Demoulin2007_overview}.

Helicity is a function of magnetic field $H=H(\bvec)$, defined as a volume integral of $\avec\cdot\bvec$ (where $\avec$ is the vector potential and $\bvec$ is the magnetic field), provided no lines of magnetic field leave this volume. For situations when field lines leave the volume (e.g., $\bvec\cdot\nhat\neq 0$ on the boundary), such as the solar corona, helicity is defined \textit{relative} to some reference field that has the same normal component of magnetic field at the boundary of the volume: $H_{relative}=H(\bvec, \bvec_{ref})=H(\bvec)-H(\bvec_{ref})$ \citep{Berger1984, Finn1985}. A potential magnetic field ($\nabla\times\bvec_{ref}=0$) has the minimum possible energy, so using it as a reference means a non-zero helicity demands some free energy. Relative helicity, defined this way, is approximately conserved under motions of plasma internal to the volume of integration.

Direct measurements of helicity in the corona remain an extremely challenging problem. They are usually performed by extrapolating magnetic field into corona using photospheric magnetic field as a boundary condition and then estimating the helicity of this field. The extrapolations are often performed in such a way that lines of resulting magnetic field resemble observed coronal loops evident in extreme ultraviolet (EUV) or soft X-Rays (SXR). The popular choices of magnetic fields are linear force-free (or constant-\al) fields (see Section~\ref{sec_method} for description) confined to a box \citep[e.g.][]{Green2002, Lim2007} and non-linear force-free field (NLFFF) extrapolations \citep{Regnier2005}. Both methods remain imperfect. The main drawbacks of the linear force-free field are that its is clearly wrong for active regions with field lines of clearly different twists \citep{Burnette2004}, and it places restrictions on \als depending on the domain size. The second method has problems dealing with solving non-linear equations and with the use of ambiguity-resolved vector magnetograms in a non force-free photosphere \citep[see][for further discussion]{Demoulin1997b}. Applying different extrapolation methods to the same data was found to produce significantly different solutions \citep{DeRosa2009}.

A quantity that is easier to measure than helicity is the flux of helicity through the photosphere. It can be shown that for changing relative helicity, changes of magnetic flux at the boundary are far more effective than internal electric currents in the presence of high connectivity \citep{Berger1984c}. This allows one to express the change in the magnetic helicity in the corona through apparent motions of photospheric magnetic features, as first suggested by \citet{Chae2001} and later developed by \citet{Demoulin2003}. 

The theoretical prediction that the coronal helicity is injected from underneath the photosphere is not supported by much observational evidence. Only a few works compare integrated helicity flux through the photosphere to coronal helicity. \citet{Pevtsov2003} have studied the evolution of coronal \als and found that its evolution is consistent with theoretical estimates of such for an emerging twisted flux tube. \citet{Burnette2004} have found a correlation between coronal \als of a linear force-free field (chosen to visually match most of SXR loops) and \als measured in the photosphere using vector magnetograms (averaged, in some sense, over the whole active region). \citet{Lim2007} have found that helicity injection through photosphere (obtained using local correlation tracking) is consistent with the change observed in the corona after taking account, approximately, of helicity carried away by CMEs; they assumes linear force-free field and a typical value for a helicity in CME. Comparisons of the change in coronal helicity with helicity of interplanetary magnetic clouds have demonstrated both a clear correspondence of the two \citep{Mandrini2004} and a lack of it \citep{Green2002}. \citet{Georgoulis2009} have compared integrated helicity flux with total helicity carried by CMEs (by multiplying a typical helicity for a CME by the amount of CMEs per studied period) and found an agreement between the two values. 

In the current work we study an emerging active region over a long period of time with high temporal cadence. We compute a rate with which a proxy for coronal helicity changes and find that this rate is consistent with rate of helicity injection through the photosphere, reported for the same active region by \citet{Longcope2007b}. This is the second time \citep[after ][]{Park2010} that the helicity \textit{change rate} in the corona over a long period of time has been found to be consistent with the helicity flux in the photosphere. 

We approximate the state of the coronal field using a method recently proposed by \citet{Malanushenko2009}. It approximates coronal loops with lines of linear force-free field like the above mentioned works, however, it allows \als to vary from line to line. That is, each of the coronal loops is approximated by a field line of a different linear force-free field. Strictly speaking, this approach is incorrect, as a superposition of linear force-free fields will not resemble a force-free field at all. \citet{Malanushenko2009} argued that it might work for some cases relevant to the solar corona and have supported the reasoning with tests on analytical non-linear (or non-constant-\al) force-free fields \citep{LowLou1990} with amount of twist comparable to one typical to solar active regions as reported by \citet{Burnette2004}. 

This method of determining properties of coronal loops does avoid some of the problems faced by linear or non-linear extrapolations. We will henceforth refer to it as (Non)-Linear Force-Free Field or shortly (N)LFFF. It introduces fewer uncertainties than a linear extrapolation, as \als is allowed to vary in space. Also it is immune to some of the drawbacks plaguing non-linear extrapolations; in particular, it does not use vector magnetograms (which are obtained in the photosphere, where the force-free approximation is questionable, see \citet{Demoulin1997b} for further discussion) and does not attempt to solve non-linear equations.

The paper is organized in the following manner. In Section~\ref{sec_method} we review the methodology, describe the measured quantity and establish its relation to helicity. In Section~\ref{sec_data} we describe the data to be used. In Section~\ref{sec_res} we summarize results and findings from the data analysis. Section~\ref{sec_discuss} contains the discussion of the results achieved. The Appendix explains tangent plane projection and why it was chosen for the present work. 
\section{The Method}
\label{sec_method}

Most active region coronal magnetic fields are believed to be in a force-free state, 
\begin{equation}
\nabla\times\bvec=\alpha\bvec, 
\label{eqn_fff}
\end{equation}
where $\alpha$ is a scalar of proportionality \citep[e.g.,][]{Nakagawa1971}. Extrapolations of coronal magnetic field usually imply solving this equation using photospheric data as a boundary condition. When $\nabla\alpha=0$,  Equation~(\ref{eqn_fff}) reduces to a system of linear equations and thus is called linear or constant-\als force-free field. When \als varies in space, Equation~(\ref{eqn_fff}) represents a non-linear system and the solution is called a non-linear of non-constant-\als force-free field. 

In the present work we use linear force-free fields confined to half space, computed using Green's function from \citet{Chiu1977}. Such fields are less popular than linear force-free fields confined to a box \citep{Alissandrakis1981} because it takes much more computational time to build them. We use them, however, because they do not impose restrictions on \als based on the size of the computational domain and because the field lines are allowed to leave the computational domain. We believe this is a better representation of coronal magnetic fields. 

The result of the procedure of the loop fitting is a set of field lines of constant-\als fields\footnote{Note that even for non-linear force-free fields $\alpha=\mbox{const}$ along every field line, of $\nabla\alpha\cdot\bvec=0$. This result is obtained by taking the divergence of both sides of Eq.~(\ref{eqn_fff}) and using $\nabla\cdot\bvec=0$ and the identity $\nabla\cdot(\nabla\times\bvec)=0$, so that $\bvec\cdot\nabla\alpha=0$ \citep{Priest_book}.}, each of them is a best fit to an individual coronal loop for every visible loop. This gives the following set of parameters for every coronal loop that was successfully fit: its \al, shape $\rvec(l)$ and the magnetic field strength along the loop $\bvec(l)$. 

Note that this method is principally different from calculating maps of $\alpha_z=J_z/B_z$ using vector magnetograms. (N)LFFF uses coronal morphology to derive currents in the corona; these could, in principle, be traced down to the upper chromosphere level and provided enough coronal loops are taken into account, a map of $\alpha_z$ could be then derived. In the previous work we have shown that such maps yield acceptable results on synthetic data. \citet{Burnette2004} have done similar analysis, but they estimated one \als that would best fit all of the observed loop of a single active region (that is, assuming the field of an active region is linear). They compared the results obtained from vector magnetograms $\langle\alpha_z\rangle$ and the results of the single-\als coronal fit $\alpha_{best}$ and reported the correlation between the two. 

In the current work we do not compute a helicity, but rather a quantity related to it, $\alpha_i L_i/2$ (where $L_i$ is the length of $i$-th field line and $\alpha_i$ is its \al). Most classical definitions of helicity involve a volume integral, which requires knowledge of magnetic field on a grid or analytically in a volume. (N)LFFF does not provide such gridded data or even volume-filling data. However, $\alpha L/2$ is closely related to helicity. For example, self-helicity of a uniformly twisted torus is $H=Tw\Phi^2$, where $Tw$ is a number of turns that a field line makes per unit length and $\Phi$ is a net magnetic flux \citep{Berger1984, Moffatt1992}, and for a thin cylindrical uniformly-twisted flux tube\footnote{Thin flux tube approximation usually refers to a structure with well-defined axis, with diameter small compared to its length, and with radius of curvature large compared to its diameter.} it can be shown that $2\pi Tw=\alpha L/2$ over axial distance $L$ (Aschwanden, 2006\nocite{Aschwanden_book}). 

It is not immediately clear that the same expression could be used for a more complex magnetic flux configuration when thin flux tube approximation is not applicable. However, \citet{Longcope2008e} have demonstrated how additive self helicity (helicity of a field relative to a potential field confined to the same domain) is consistent with an empirical function $\alpha \langle L_i\rangle\Phi^2/4\pi$ with $\langle L_i\rangle$ being the average length of a field line in the domain\footnote{Here ``domain'' is a volume occupied by field lines connecting two given footpoints, so there is no magnetic flux across the boundaries of such a volume, except at the footpoints.} and $\Phi$ being the total magnetic flux in the domain. This was shown for a case when the thin flux tube approximation was clearly inapplicable: for a linear force-free field of a quadrupolar field confined to a box, whose domain is the field connecting two polarities. Later \citet{Malanushenko2009a} suggested that ``flux-normalized'' additive self helicity could be treated as a generalized twist, $H_A/\Phi^2=Tw_{gen}$ for an arbitrary magnetic configuration. That leads to a conclusion that 
\begin{equation}
2\pi Tw_{gen}=\alpha \langle L_i\rangle/2,
\label{eqn_tw_gen_linear}
\end{equation}
in a similar manner as in thin flux tubes, at least in \textit{linear} force-free fields. \citet{Malanushenko2009a} demonstrated that $Tw_{gen}$ is equal to $Tw$ for a thin flux tube and that it behaves like $Tw$ in arbitrary magnetic configurations, for example, serving as a kink instability threshold. We hereafter refer to $Tw_{gen}=H_A/\Phi^2$ as ``twist''.
 
To the best of our knowledge, such relationship between \als and helicity has not been established for general non-linear force-free fields. However in a thin \textit{non-}uniformly twisted cylinder it might be expected that $Tw_{gen}$ would be proportional to axial length with a constant of proportionality, that should reduce to \als if the cylinder were uniformly twisted. For a more complex configuration $Tw_{gen}$ might be expected to be proportional to some length scale times a constant that has dimensions of \al. We choose to use a quantity $\langle\alpha_i L_i/2\rangle$, averaged over many field lines as a proxy of $Tw_{gen}$ and find that it changes consistently with the injection of coronal twist. 
 
\begin{figure}[!hc]
 \begin{center}
  \begin{tabular}{p{5.0cm}p{5.0cm}}
   \includegraphics[height=5cm]{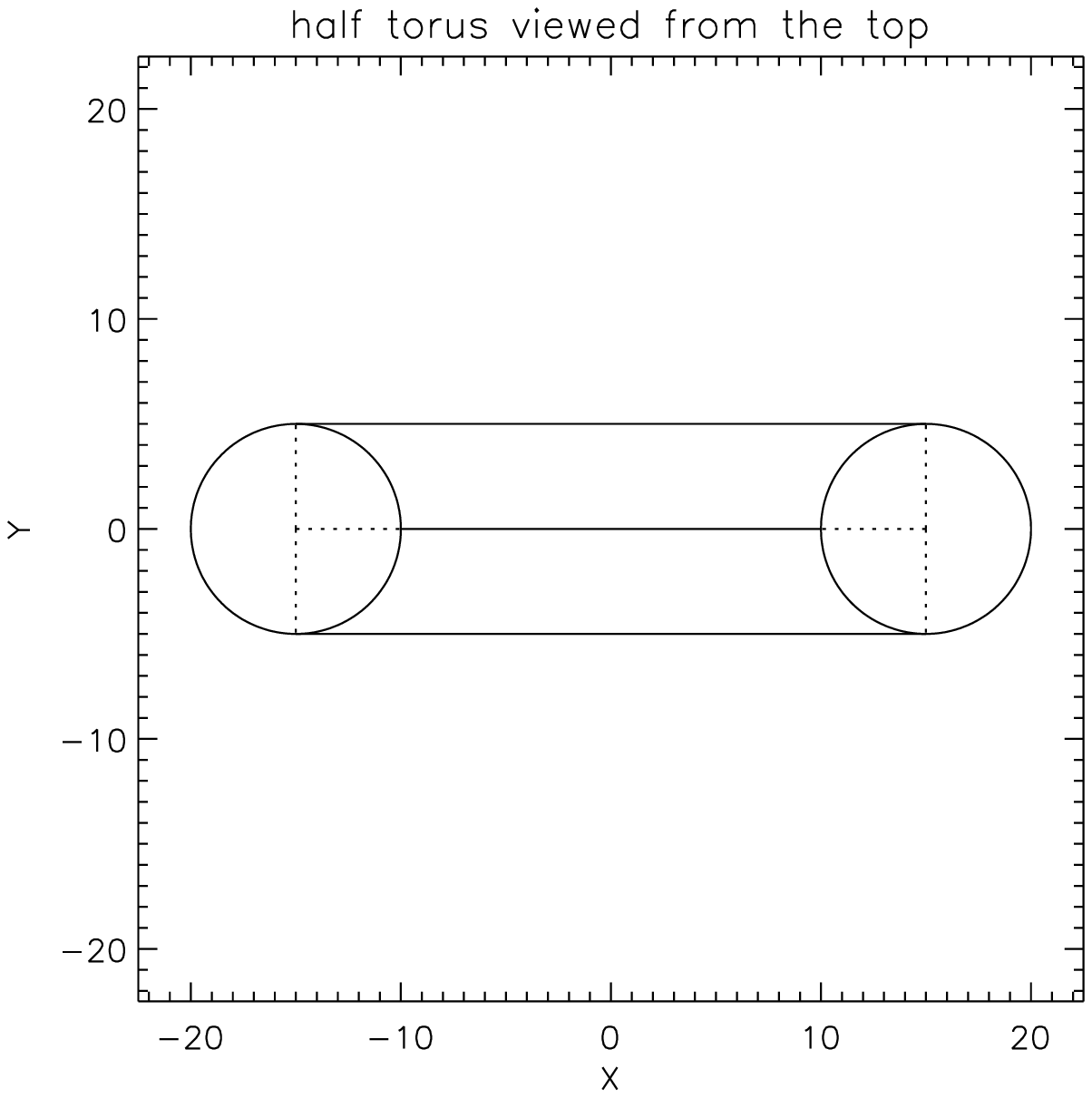} & 
    \includegraphics[height=4cm]{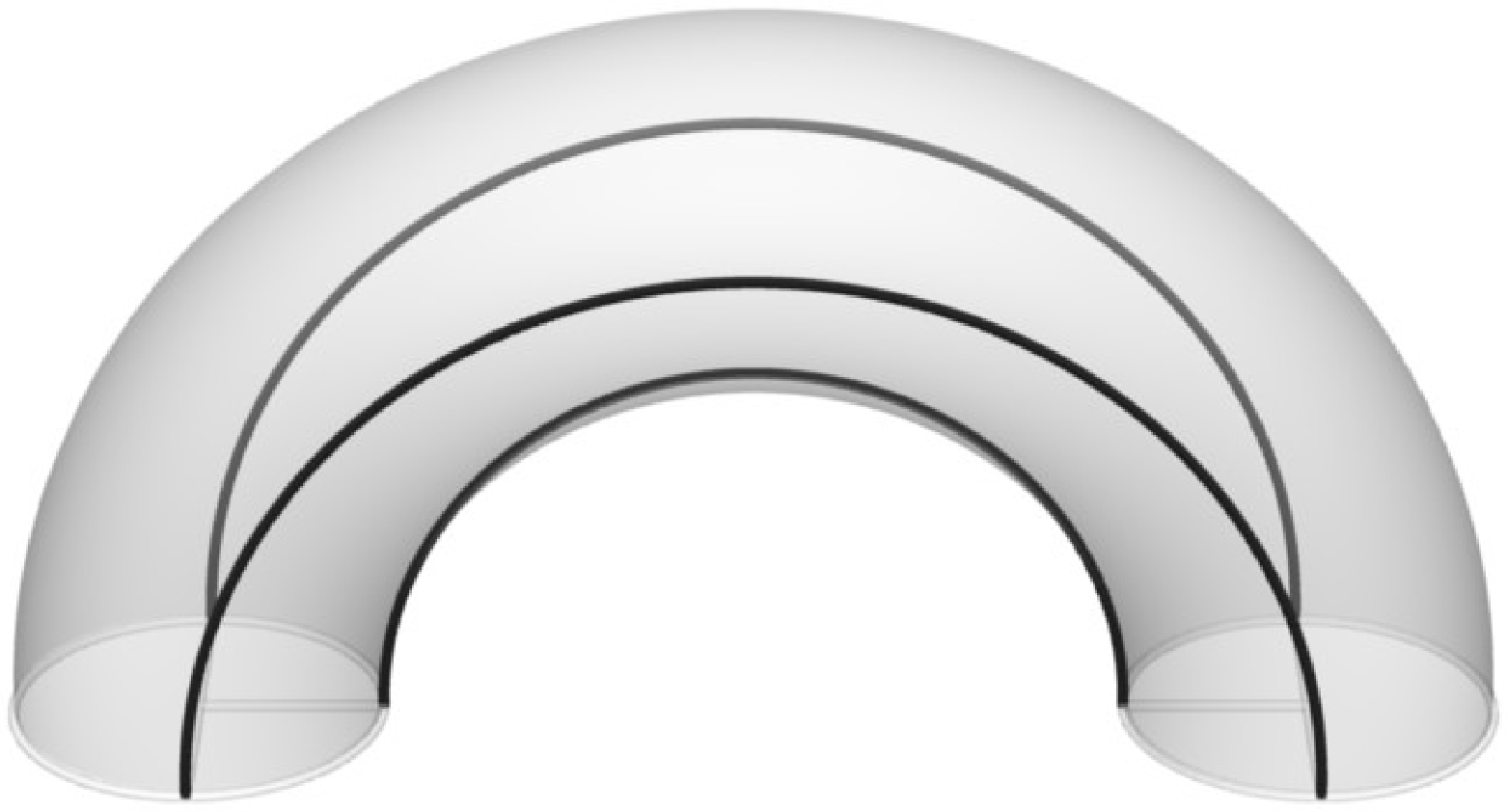} \\
   \includegraphics[height=5cm]{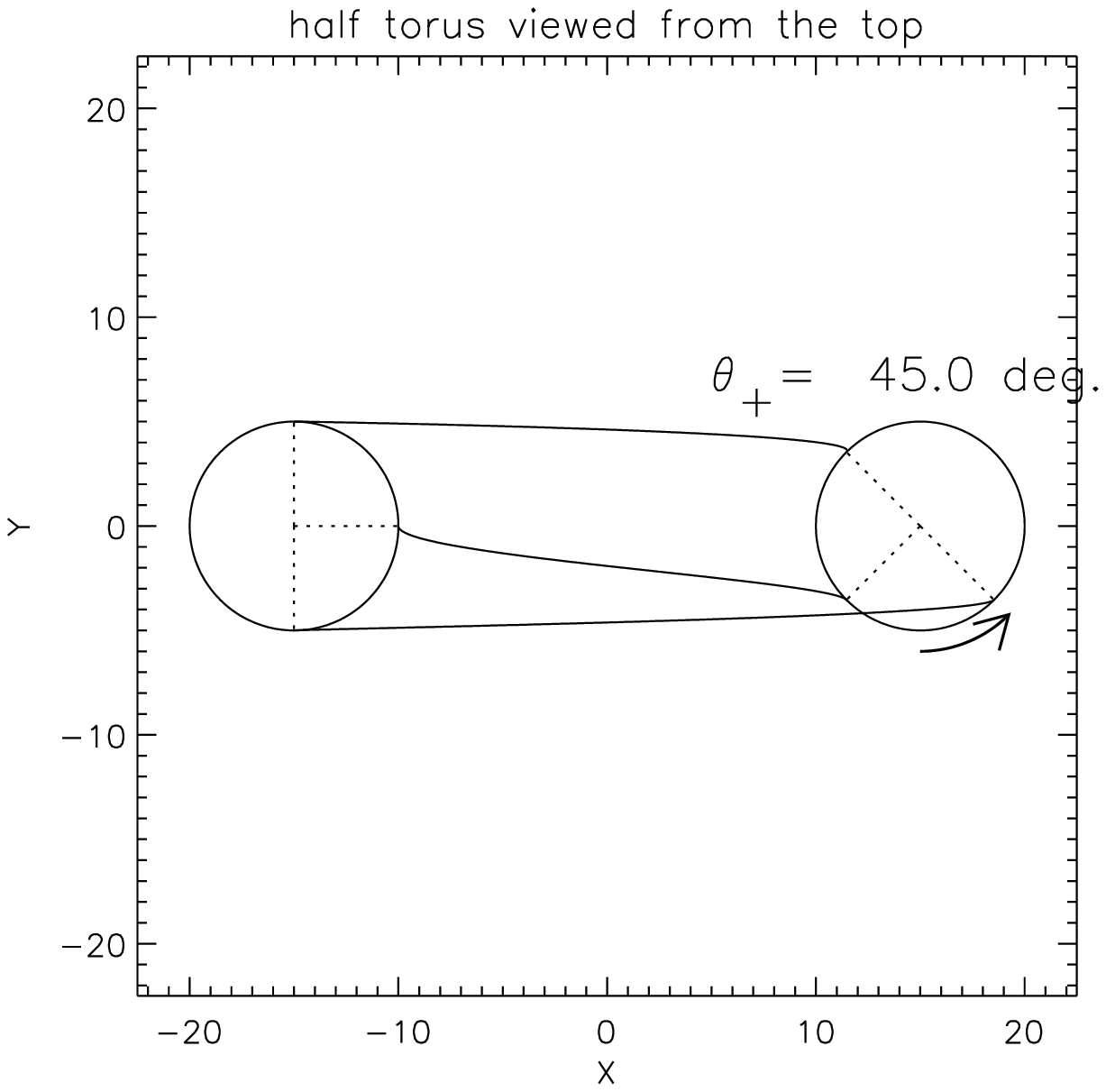} & 
    \includegraphics[height=4cm]{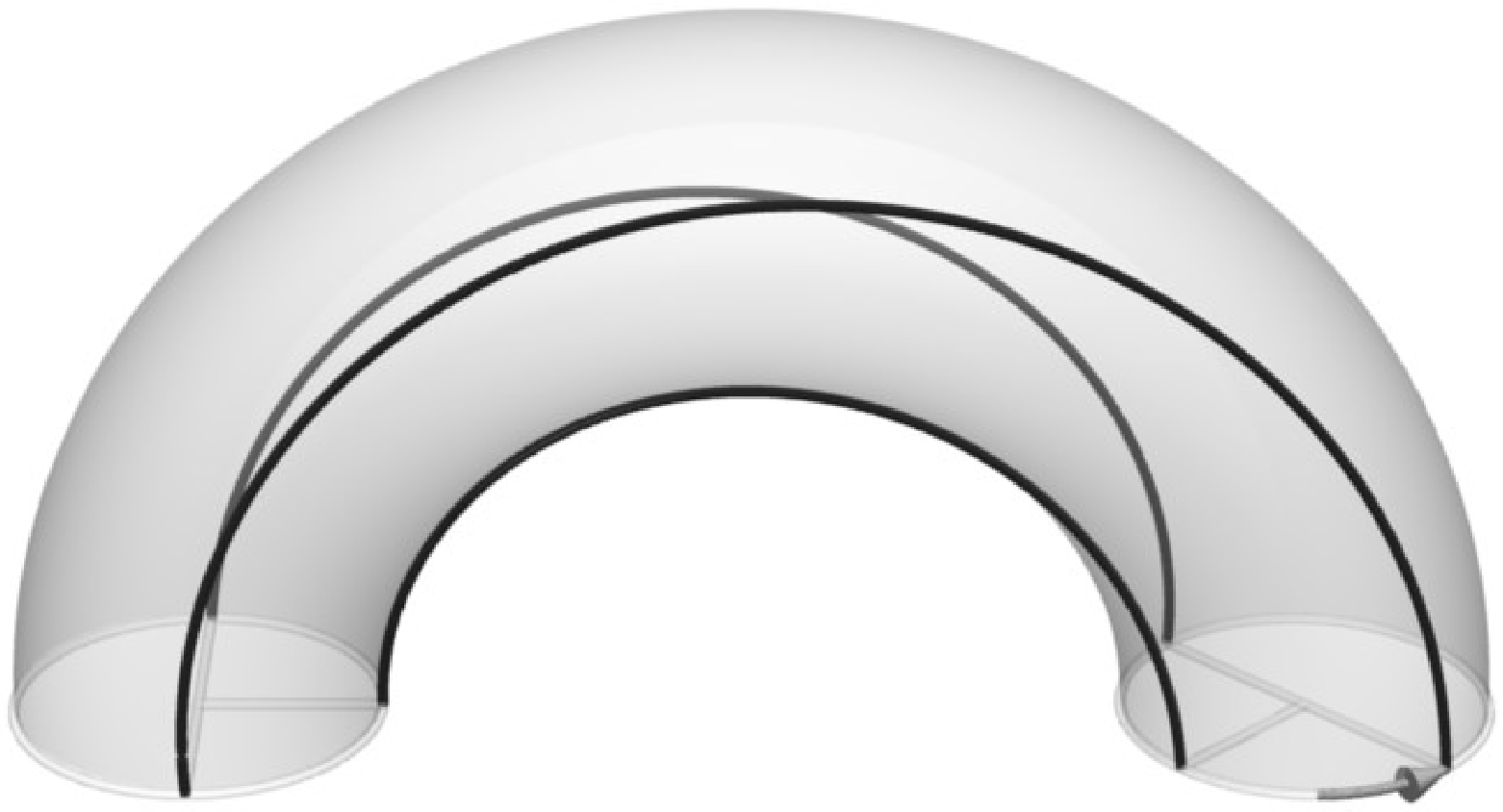} \\
   \includegraphics[height=5cm]{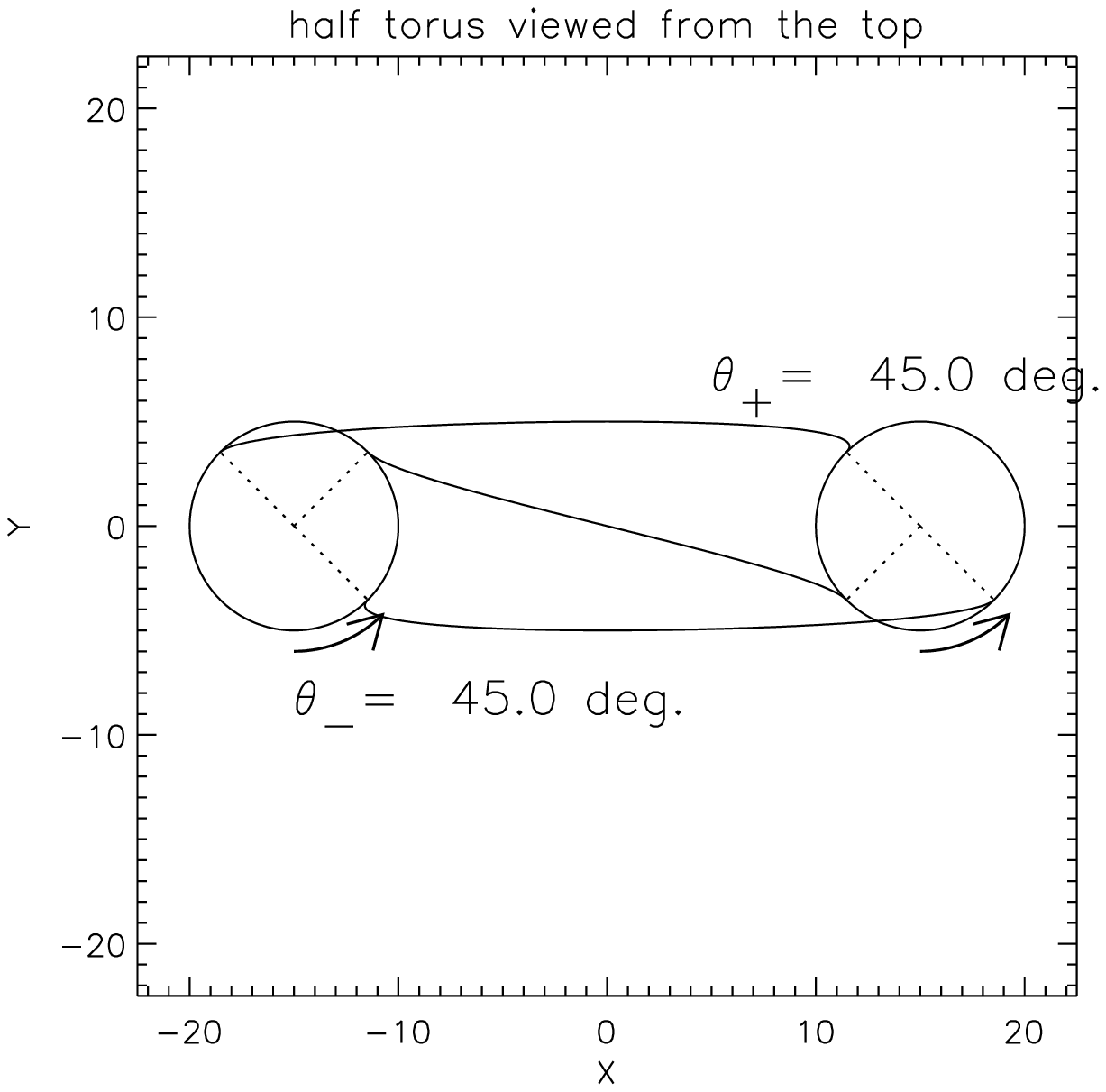} & 
    \includegraphics[height=4cm]{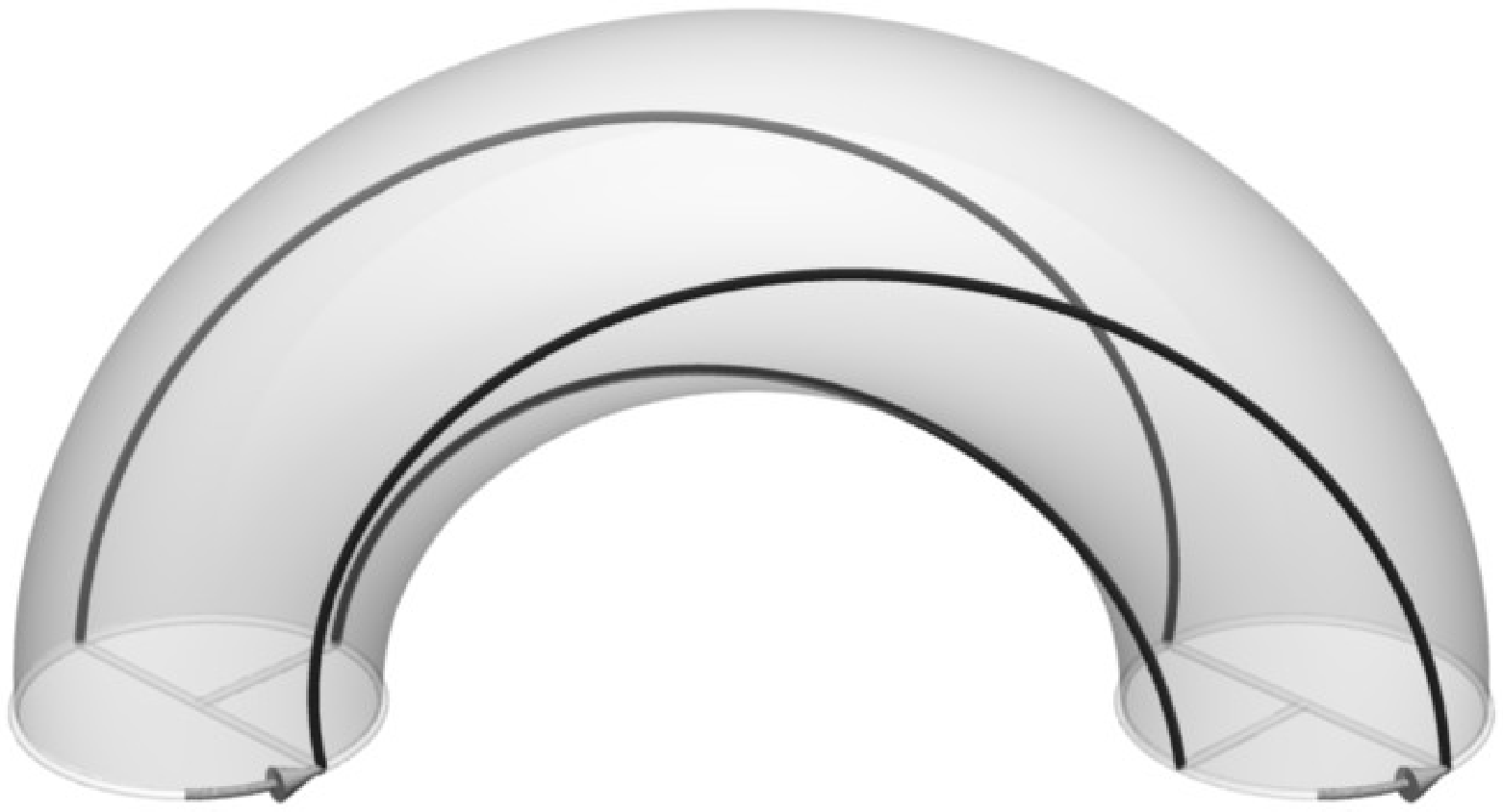} \\
   \includegraphics[height=5cm]{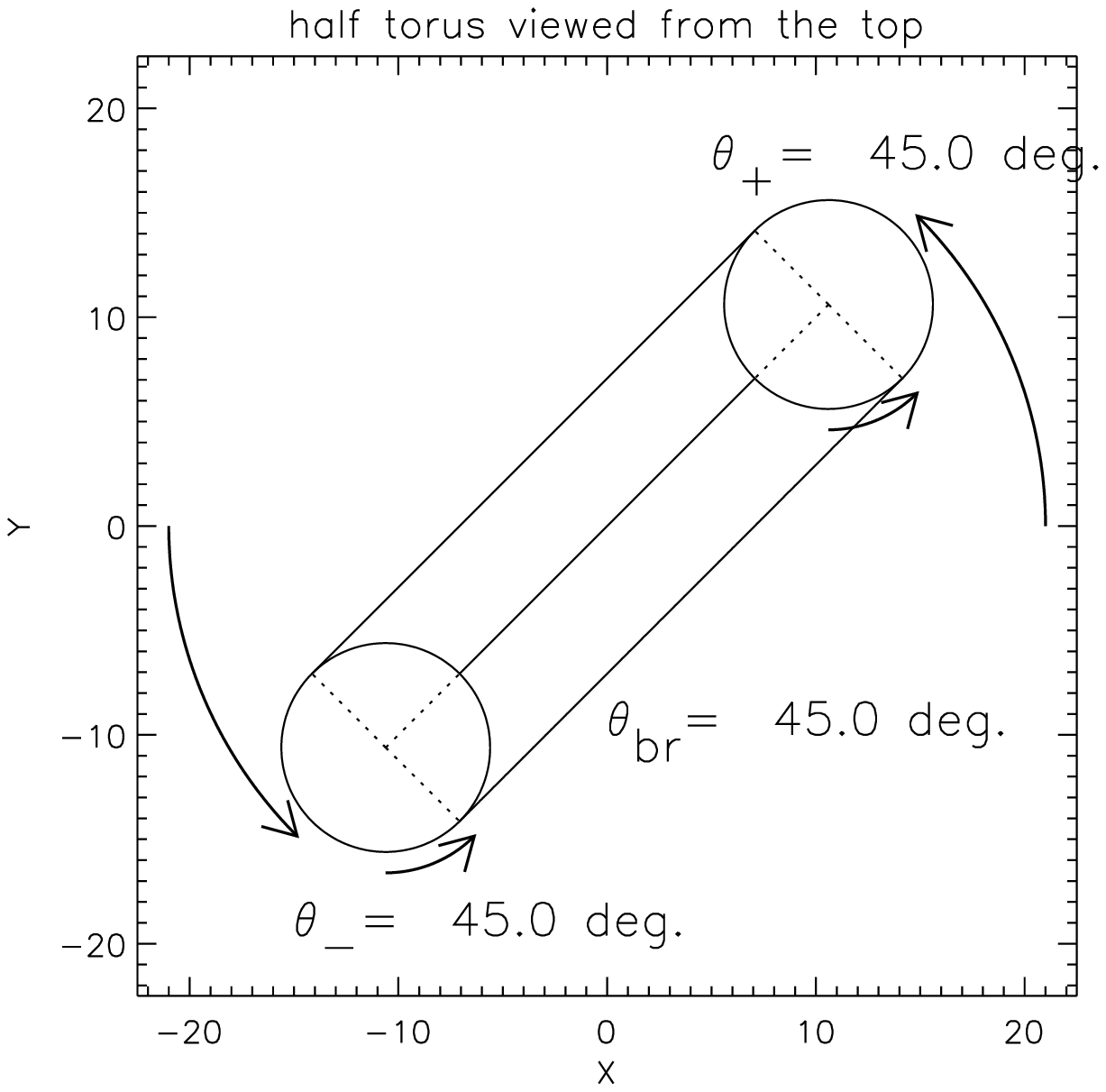} & 
    \includegraphics[height=4cm]{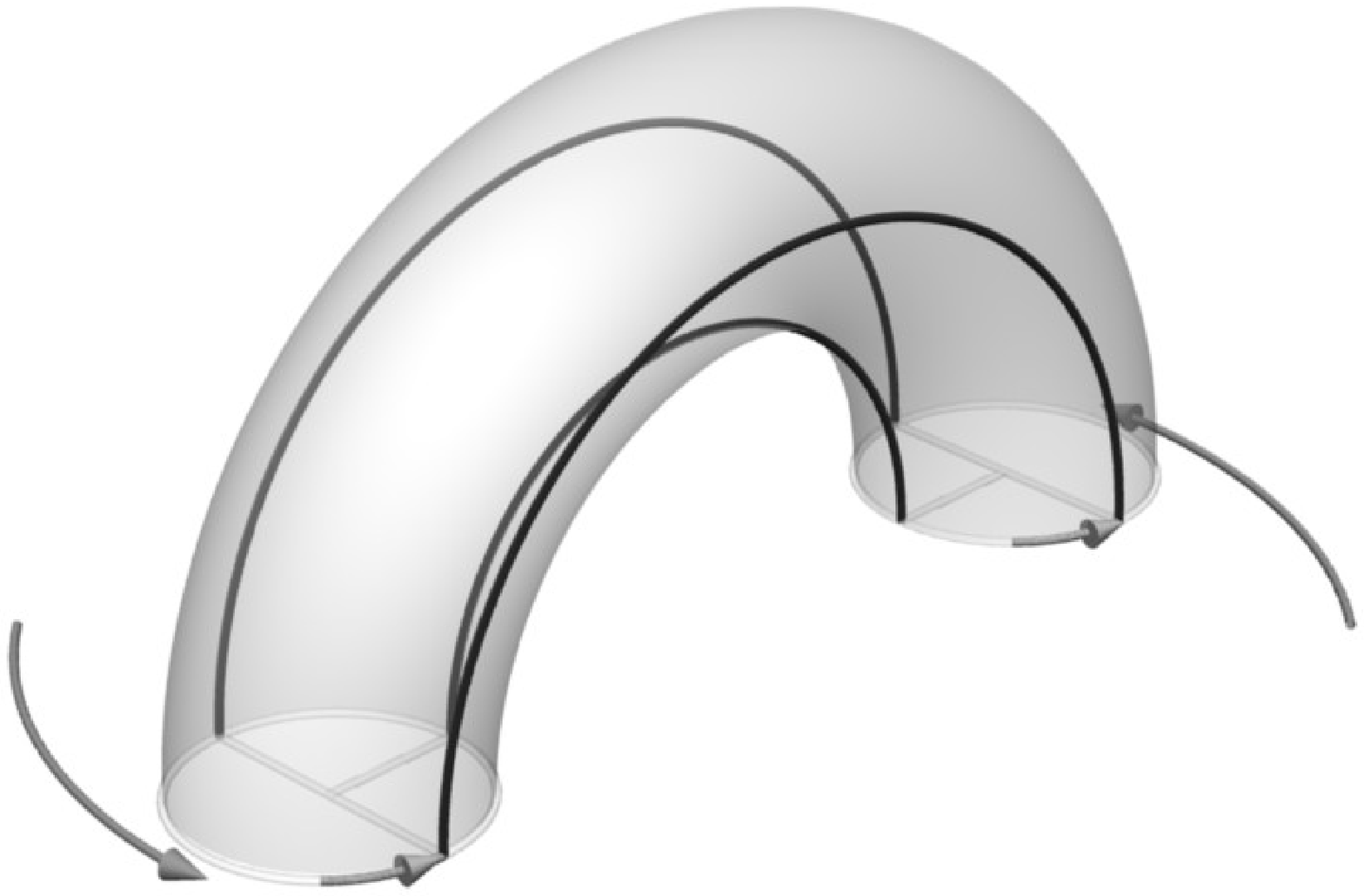} \\
  \end{tabular}
 \end{center}
 \caption{\small{A schematic illustration of how braiding and spinning motions of footpoints contribute to helicity. \textit{(top row)} An untwisted half torus. \textit{(two rows in the middle)} Each of the footpoints is rotated about its own center by angle $\theta_+=\theta_-=\theta$ (spinning motion). \textit{(bottom row)} If then the footpoints are rotated about each other (braiding) by $\theta_{br}=\theta$, the resulting configuration would have no helicity in it, in agreement with Equation~\ref{eqn_hel_total}.}}
 \label{torus}
 \end{figure}

We compare this quantity to the normalized helicity flux across the photosphere, measured by \citet{Longcope2007b}. They follow the results of \citet{Welsch2003}, who start with the commonly accepted expression for helicity flux \citep{Berger1984}
\begin{equation}
\frac{dH}{dt}=2\int\limits_{S}{[(\avec_P\cdot\bvec)\vvec-(\avec_P\cdot\vvec)\bvec]\cdot\nhat dS}, 
\label{h_flux_welsch2003}
\end{equation}
here $\nhat$ is a normal vector pointing into the volume and $\avec_P$ is assumed to be in Coulomb gauge with  $\avec_P\cdot\nhat=0$. They split it into two terms, that depend on the normal and tangential components of the velocity on the boundary. These two terms correspond to the change in helicity due to the emergence of the new magnetic flux and to the surface motions of the existing magnetic flux respectively. Assuming the boundary is $z=0$ plane, the second term can then be written as 
\begin{equation}
\frac{dH}{dt}\left(\vvec_\perp\right)=2\int\limits_{S}{(\avec_P\cdot\vvec_\perp)B_zdS}.
\label{h_flux_welsch2003_part}
\end{equation}
\citet{Welsch2003} further concentrate on this term only, thus discarding the helicity flux due to change in the magnetic flux. They split the flux of unconfined self helicity (of a field confined to half space relative to the potential field in half space) into two parts: the ``spinning'' and the ``braiding'' contributions, as illustrated on Figure~\ref{torus}. The first one comes from rotation of footpoints about their axis and its change rate was expressed by \citet{Longcope2007b} through the average spinning rate $\dot{\theta}_{spin}$ as: 
\begin{equation}
\dot{H}_{spin}^a=-\frac{1}{2\pi}\Phi_a^2\dot{\theta}_{spin}^{a},
\label{eqn_spin}
\end{equation}
for $a$-th footpoint. The second one comes from the relative rotation of footpoints and its change rate could be expressed through the average tilt angle $\theta_{braiding}$: 
\begin{equation}
\dot{H}_{braiding}^{ab}=-\frac{1}{\pi}\Phi_a\Phi_b\dot{\theta}_{braiding}^{ab}.
\label{eqn_braid}
\end{equation}

$\dot{H}_{spin}$ and $\dot{H}_{braiding}$ are not equivalent to fluxes of ``twist'' and ``writhe'' helicities \citep{Berger1984, Moffatt1992}. In addition to the discussions in the previously mentioned papers, we would like to provide an illustrative example of this statement. Consider a thin untwisted half torus from Figure~\ref{torus} (top row), deformed in the following way: first, each of the footpoints is rotated about its own center by $\theta$ (second and third rows), then the footpoints are rotated about each other by $\theta$ (bottom row). This transformation is equivalent to a rotation of the whole torus as a rigid body and adds no inner twist to it. However, an observer, who believes that $\dot{H}_{braiding}$ is equivalent to the flux of writhe helicity, might consider that the writhe helicity has increased by $\Phi^2\theta/\pi$, while the final configuration is the same as the starting configuration in a rotated reference frame. \citet{Longcope2007b} argued that ``spinning'' and ``braiding'' fluxes of helicity might be produced by different physical processes.
 
Combining Equations~(\ref{eqn_spin}) and~(\ref{eqn_braid}) and noticing that for a magnetic flux balanced dipole $\Phi_a=-\Phi_b$, we get: 
\begin{equation}
\dot{H}_{total}/\Phi^2=-\frac{1}{2\pi}\left(\dot{\theta}_{spin}^{a}+\dot{\theta}_{spin}^{b}-2\dot{\theta}_{braiding}^{ab}\right).
\label{eqn_hel_total}
\end{equation}
From Equations~(\ref{eqn_tw_gen_linear}) and~(\ref{eqn_hel_total}) it follows, that if the quantity $\langle\alpha_i L_i\rangle/2$ is roughly proportional to $H_{A}/\Phi^2$ in a non-linear force-free field, \begin{equation}
H_{A}/\Phi^2\approx\frac{1}{2\pi}\langle\alpha_i L_i\rangle/2,
\label{eqn_hel_main}
\end{equation}
its change rate should be proportional to $-\left(\dot{\theta}_{spin}^{a}+\dot{\theta}_{spin}^{b}-2\dot{\theta}_{braiding}^{ab}\right)$. 

Note that Equation~(\ref{eqn_hel_total}) only represents the term given by Equation~(\ref{h_flux_welsch2003_part}), which assumes constant magnetic flux. Since we are interested in comparing results of the photospheric flux of helicity with ours, we follow the same reasoning and neglect the effects due to the change in magnetic flux as well. It was pointed out by \citet{Demoulin2003} that the horizontal motion of flux tubes would, in principle, include both contributions to Equation (\ref{eqn_hel_total}), including that from vertical flow (emergence). It therefore follows that if tracking methods used to measure horizontal speeds tracked magnetic footpoints, then their use in Equation (\ref{eqn_hel_total}) would also capture both contributions.  Tests of these tracking methods, however, casts some doubt on this premise \citep{Welsch2007}. Local Correlation Tracking seems to capture the horizontal flow speed far more accurately than it captures the contribution of vertical flow.  Its application to Equation (\ref{eqn_hel_total}) thus captures most reliably the first term, neglecting, or underestimating, the contribution of emergence.

$\langle\alpha_i L_i\rangle/2$ is really meant to represent $Tw_{gen}$ and thus be relevant only to the additive self-helicity $H_A$, that is to helicity of the field in a subdomain relative to the potential field in the same subdomain, and $H_{total}$ is the helicity of the field in half space relative to potential field in half space. However, the difference $H_{total}-H_A$ only depends on the shape of the subdomain and both \citet{Longcope2008e} and \citet{Malanushenko2009a} have studied the cases with subdomains of relevant shapes and found this difference to be small (except at the pre-eruption state). 

\section{The Data}
\label{sec_data}

 \begin{figure}[!hc]
 \begin{center}
  \begin{tabular}{p{7.0cm}p{7.0cm}}
   \includegraphics[height=7cm]{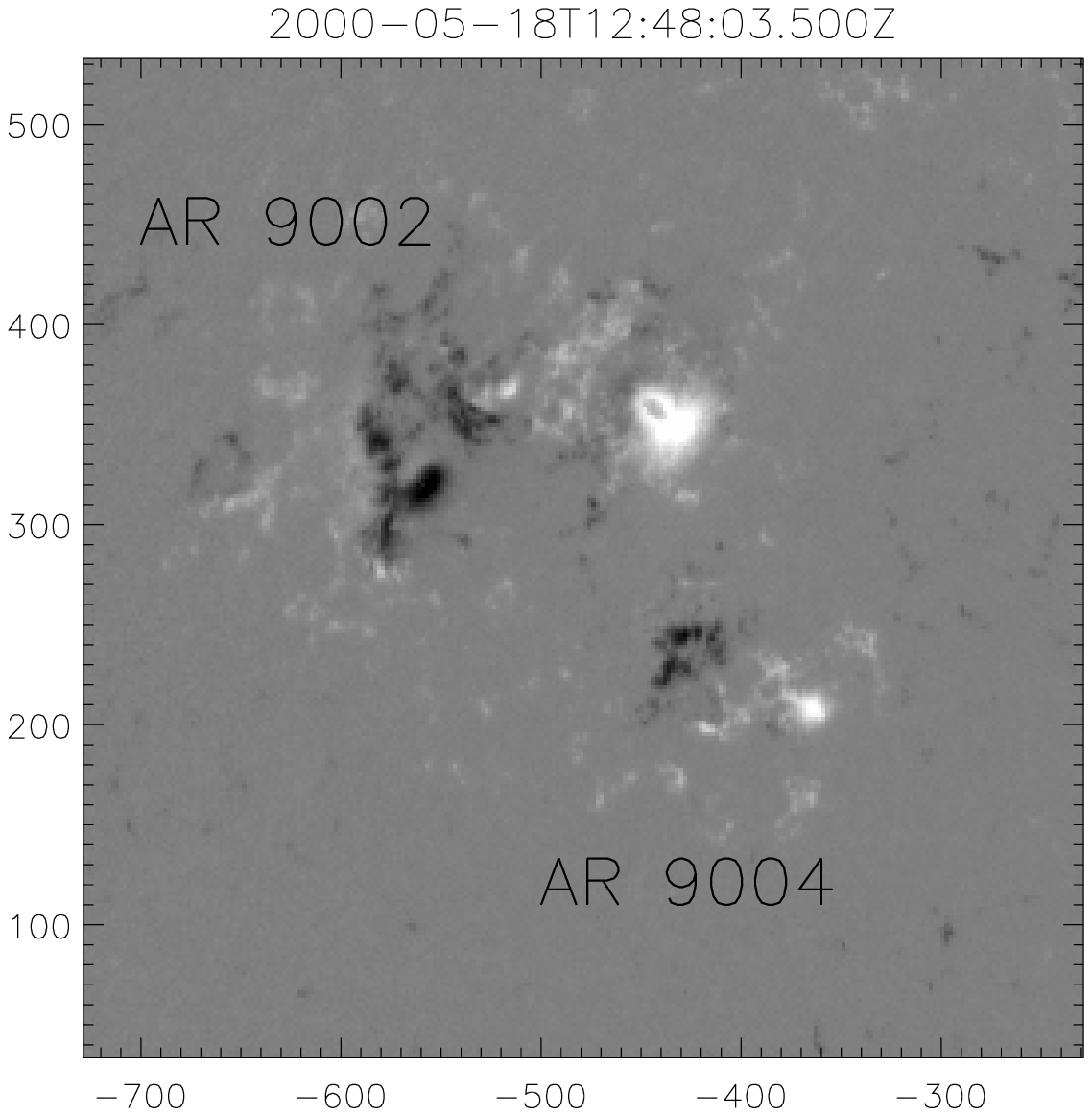} &
   \includegraphics[height=7cm]{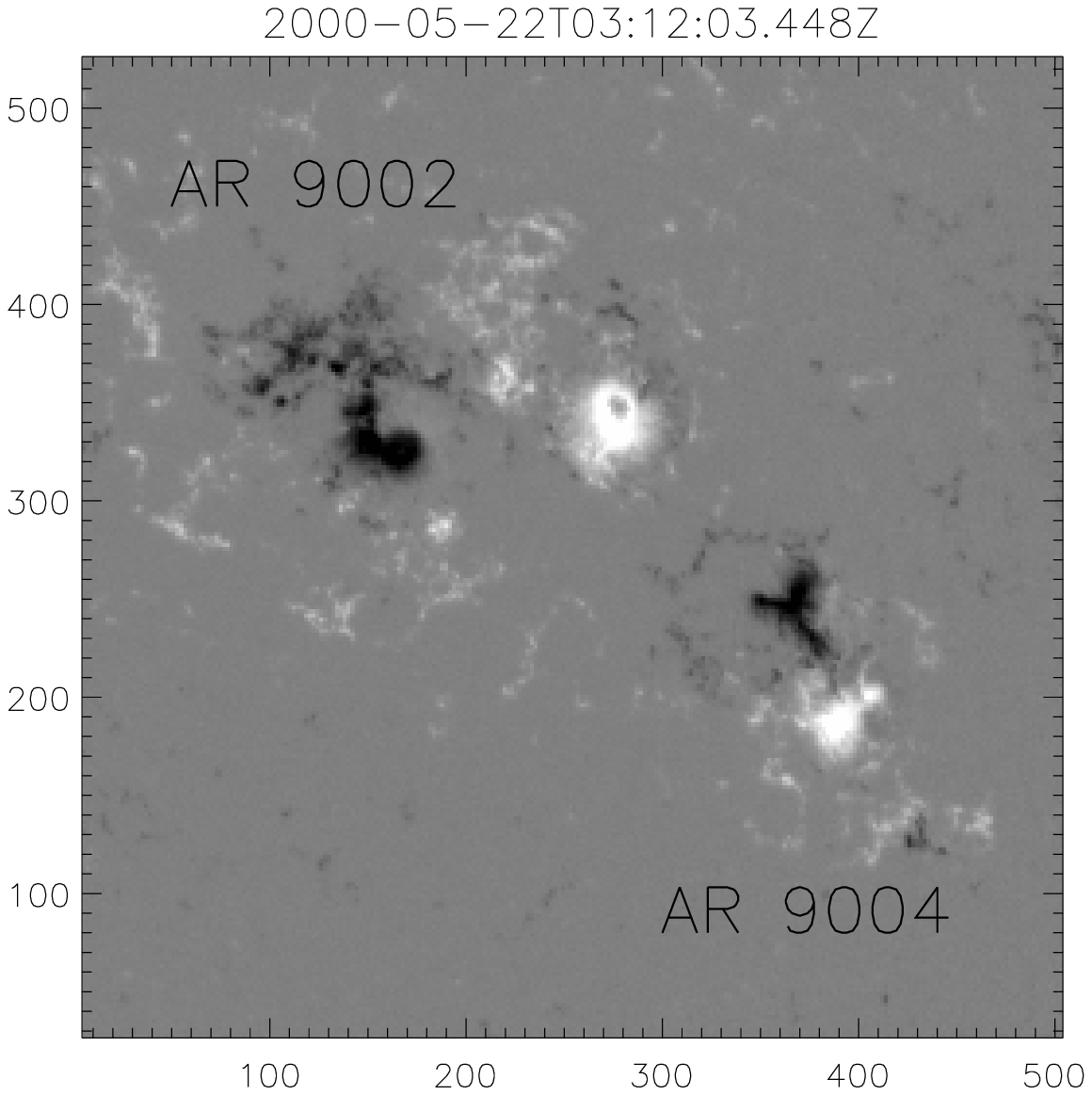} \\
  \end{tabular}
 \end{center}
 \caption{\small{We studied the region with AR's 9002 and 9004 from 2000-05-18 12:48:03 till 2000-05-22 03:12:03. AR 9002 was an old diffuse active region and AR 9004 was emerging and rapidly rotating active region.}}
 \label{data_intro}
 \end{figure}

The present study is devoted to AR 9004. It was emerging, as its magnetic flux was increasing. At the earliest time when the magnetograms were available it had about 30\% of its final magnetic flux. The other feature of this active region is that its footpoints were rotating about each other and about themselves at measurable pace. 

While we have only analyzed AR 9004, we included AR 9002 in the study. There are several reasons for this, such as proximity of the two active regions and the appearance of coronal loops interconnecting them. Also, while many loops could definitely be attributed only to one active region, to the other one, or connecting the two, the connectivity of other loops was not so clear. We included both ARs in the computational domain and reconstructed all coronal loops in TRACE 171\AA~field of view (that had both active regions). For the further analysis we only considered the loops that were found to start and end at AR 9004. 

We chose 21 full-disk MDI magnetograms in the time range between 2000-05-18 12:48 and 2000-05-22 03:12. The first and the last images are shown on Figure~\ref{data_intro}. 

The magnitude of the magnetic field on each magnetogram was corrected for the line-of-sight factor, assuming the magnetic field to be purely radial: $B_z(x, y)=B_{los}(x, y)/\sqrt{1-\rho(x, y)^2/R_{\Sun}^2}$, where $\rho(x, y)=\sqrt{x^2+y^2}$ is the plane-of-the-sky distance from the disk center to the point $(x, y)$. We did not account for possible inclination of the magnetic field in active regions \citep{Howard1991_7}. The largest average inclination angle \citet{Howard1991_7} found for growing ARs to be $25^{\circ}$ for the leading polarity. The error in the strength of the photospheric magnetic field that we make assuming the field is purely radial is thus about $\cos 25^{\circ}\approx 0.9$. It is not clear how big of a difference would it make, but in Malanushenko et. al. (2009) we report uncertainties of other origin that are bigger than 10\%. We acknowledge that the analysis in the present work has multiple sources of uncertainties, mentioned in this paragraph and further through the text; they could have a cumulative effect resulting in the real uncertainty bigger than 10\%. 

In the next step, each magnetogram was then remapped to the disk center in orthographic projection (see Appendix). This was done to correct for the foreshortening distortions of the magnetogram. This step was necessary as during the observed time sequence the center of the couple 9002/9004 has traveled approximately from 15$^{\circ}$N30$^{\circ}$E to 15$^{\circ}$N16$^{\circ}$W. 
 
For each magnetogram we selected several (typically two) TRACE 171\AA~images taken within 30 minutes of the magnetogram. The selection criteria was the visibility of many distinct coronal loops. On each TRACE image as many loops as could be discerned were ``traced'', or visually approximated with a smooth curve, a two-segment B\'ezier spline \citep[e.g., ][]{BezierRef}. 

Each magnetogram on the tangent plane was rebinned on a coarser grid, half the size in each dimension, and than used to construct 41 linear force-free fields with $\alpha\in[-0.05, 0.05]\mbox{ arcsec}^{-1}$ and with step of $\Delta\alpha=0.0025\mbox{ arcsec}^{-1}$ in a box with $z_{max}=200\mbox{ arcsec}$ and $\Delta z\approx1.5\mbox{ arcsec}$. We used a threshold of 100G, that is, we did not account for weaker magnetic fields. The reason for that was to save computational time. The objects of interest were coronal loops of the scale of an active region, and such threshold is not likely to introduce significant disturbance in them. Both ranges of \als and $z$ have proven to be sufficient to reconstruct most of the loops. 

\begin{figure}[!hc]
 \begin{center}
    \vspace{0.5cm}
    \begin{tabular}{p{9.0cm}p{7.0cm}}
     \includegraphics[height=6cm]{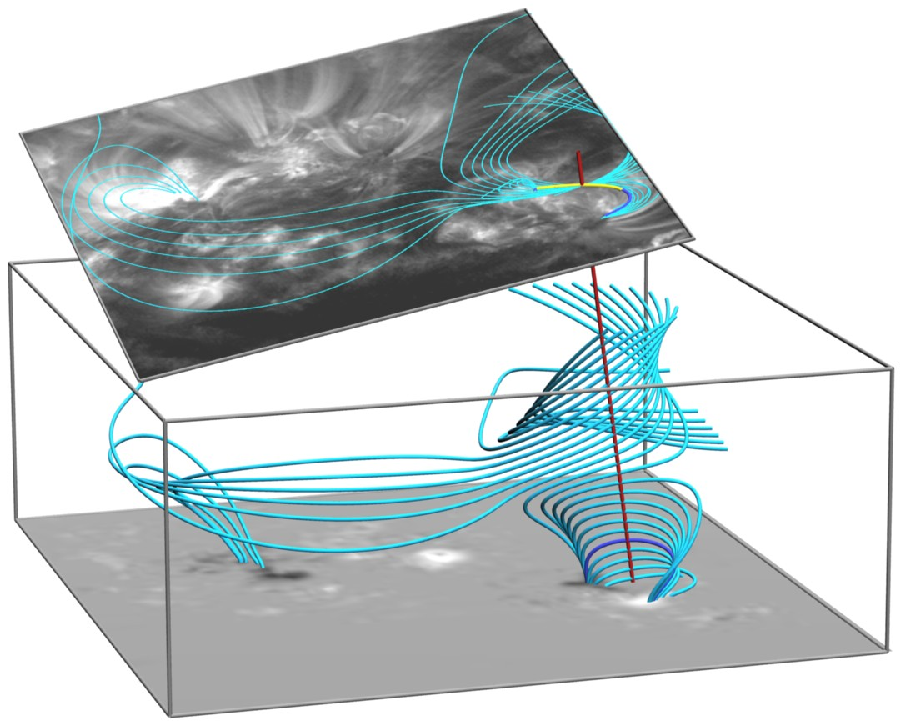} &
     \includegraphics[height=6cm]{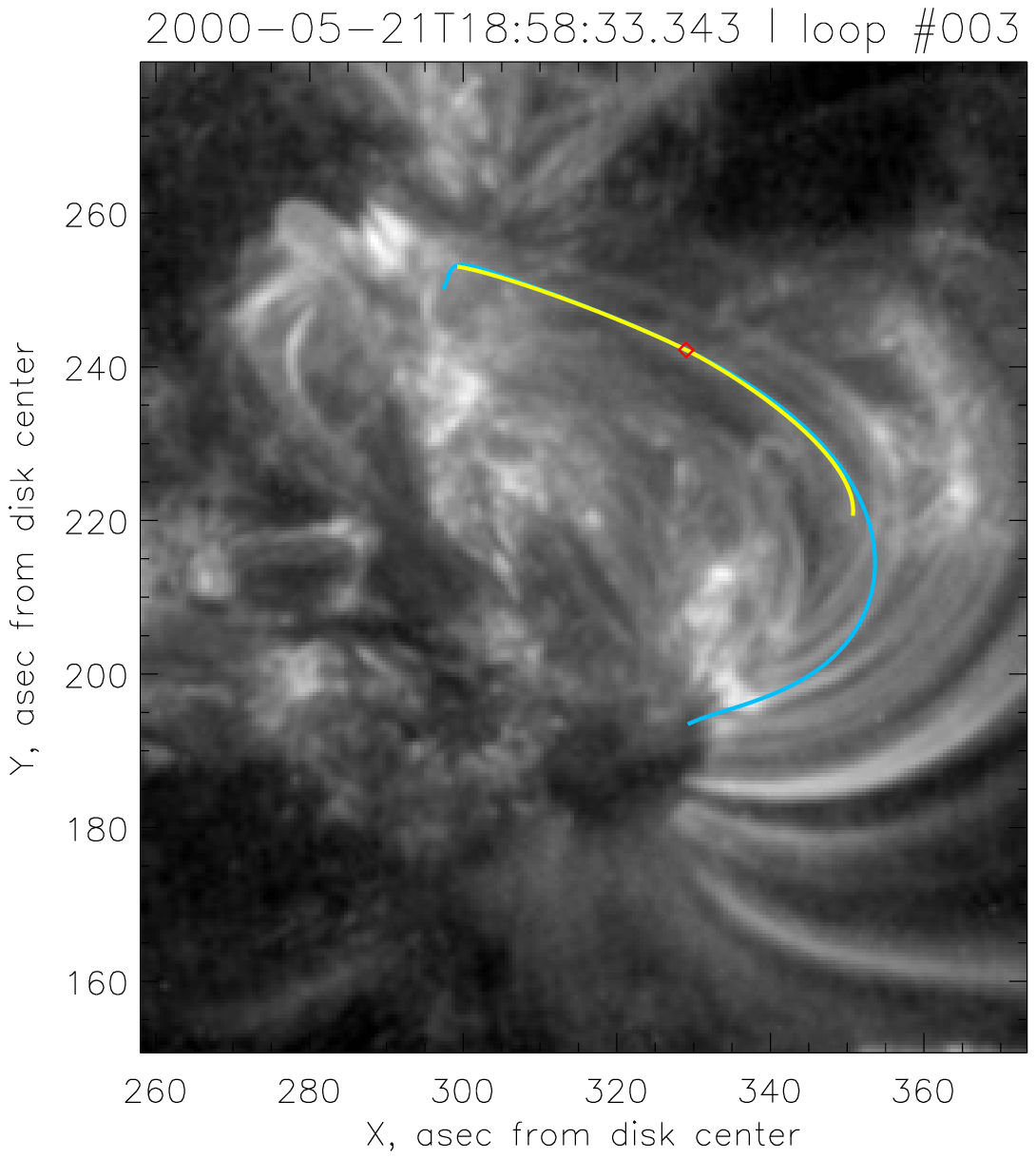} \\
    \end{tabular}
 \end{center} 
 \caption{\small{\textit{(left)} -- TRACE image in the plane of the sky and de-rotated MDI images shown with their respective angles. For every coronal loop (yellow) the algorithm browses through several different linear force-free fields. In each field it traces field lines (cyan) along the line or sight (red). The field lines are than projected back to TRACE image and compared to the loop using the semi-automatic selecting algorithm. Best-fit field line is shown in darker shade of blue. (Field lines are of $\alpha=0.025\mbox{ arcsec}^{-1}$, that corresponds to the best-fit \als for this particular loop, so the dark blue line is the resulting best-fit line for this loop.) Gray box shows the actual computational domain. \textit{(right)} -- a fragment of the TRACE image in the plane of the sky, the traced loop (yellow), the projection of the best fit to the plane of the sky (cyan) and the midpoint of the loop (red).}}
  \label{fig_3d} 
 \end{figure}

For every ``traced'' coronal loop and for many linear force-free fields we have computed many field lines along the line of sight that cross the midpoint of the loop. Figure~\ref{fig_3d} illustrates such field lines for one of the linear force-free fields. All of these field lines were discarded except for the ``best-fit'', shown on Figure~\ref{fig_3d} (right) in yellow. The selection procedure was the following: each field line (that is defined by \als of the linear force-free field it belongs to and the coordinate along the line of sight $h$) was projected onto the plane of the sky. Then the mean distance $d(h, \alpha)$ was computed between this projection and the traced loop. Then the minimum of $d(h, \alpha)$ on a grid of given values \als and $h$ was found following a semi-automatic algorithm described in \citet{Malanushenko2009}. 

We found that for many loops ($\sim$55\% of all 301 loops) the general shape of the $d(h, \alpha)$ parameter space matches one of the types described in the fitting scheme. As for the ``undescribed'' types, we chose not to ignore them, but try to identify the non-hyperbolic valley and choose a local minimum on it that resulted in visually better fit. 

\begin{figure}[!hc]
 \begin{center}
  \begin{tabular}{m{8cm}m{8cm}}
   \multicolumn{2}{c}{(a)\includegraphics[width=9cm]{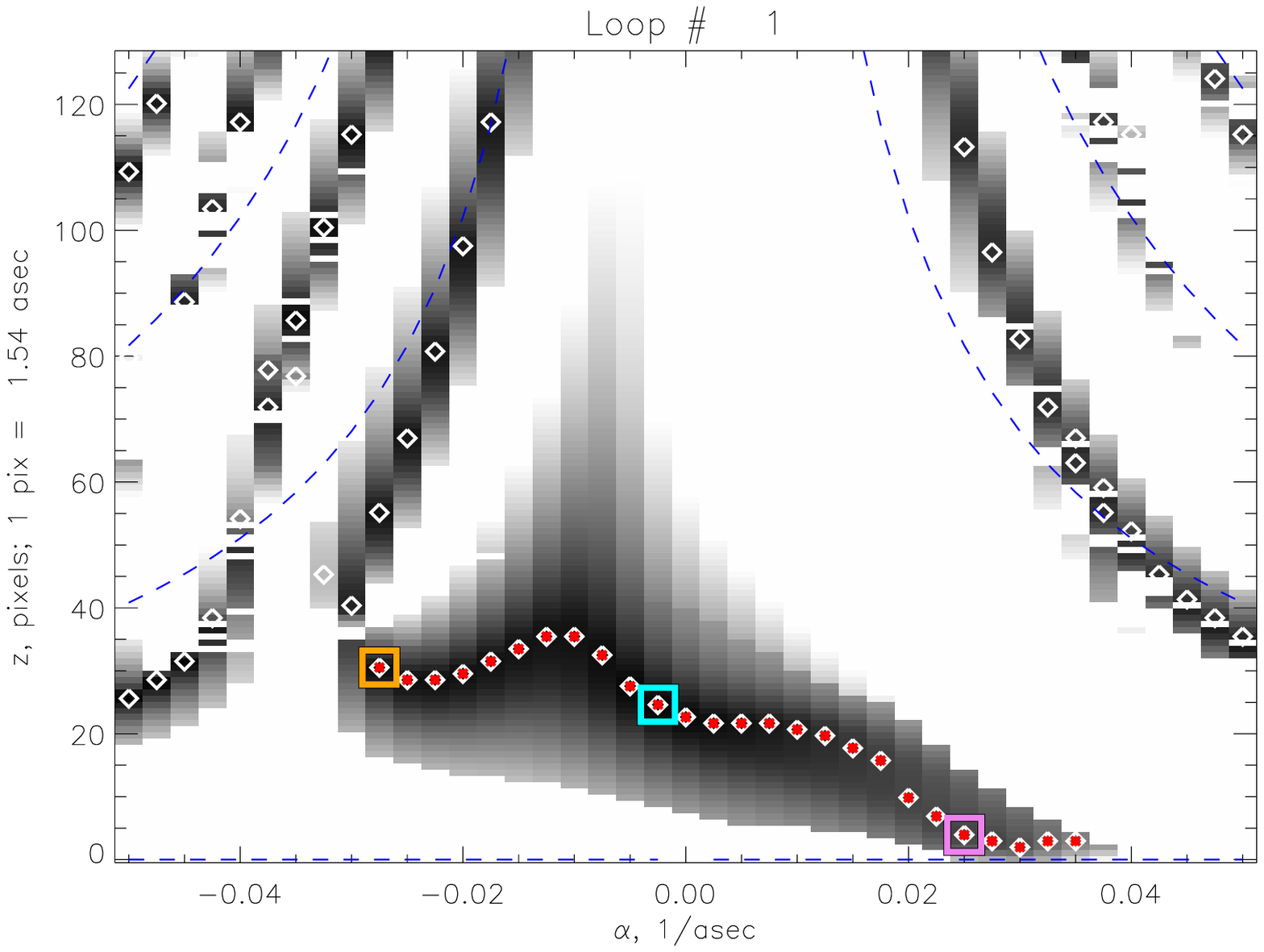}}\\
   (b)\includegraphics[width=8cm]{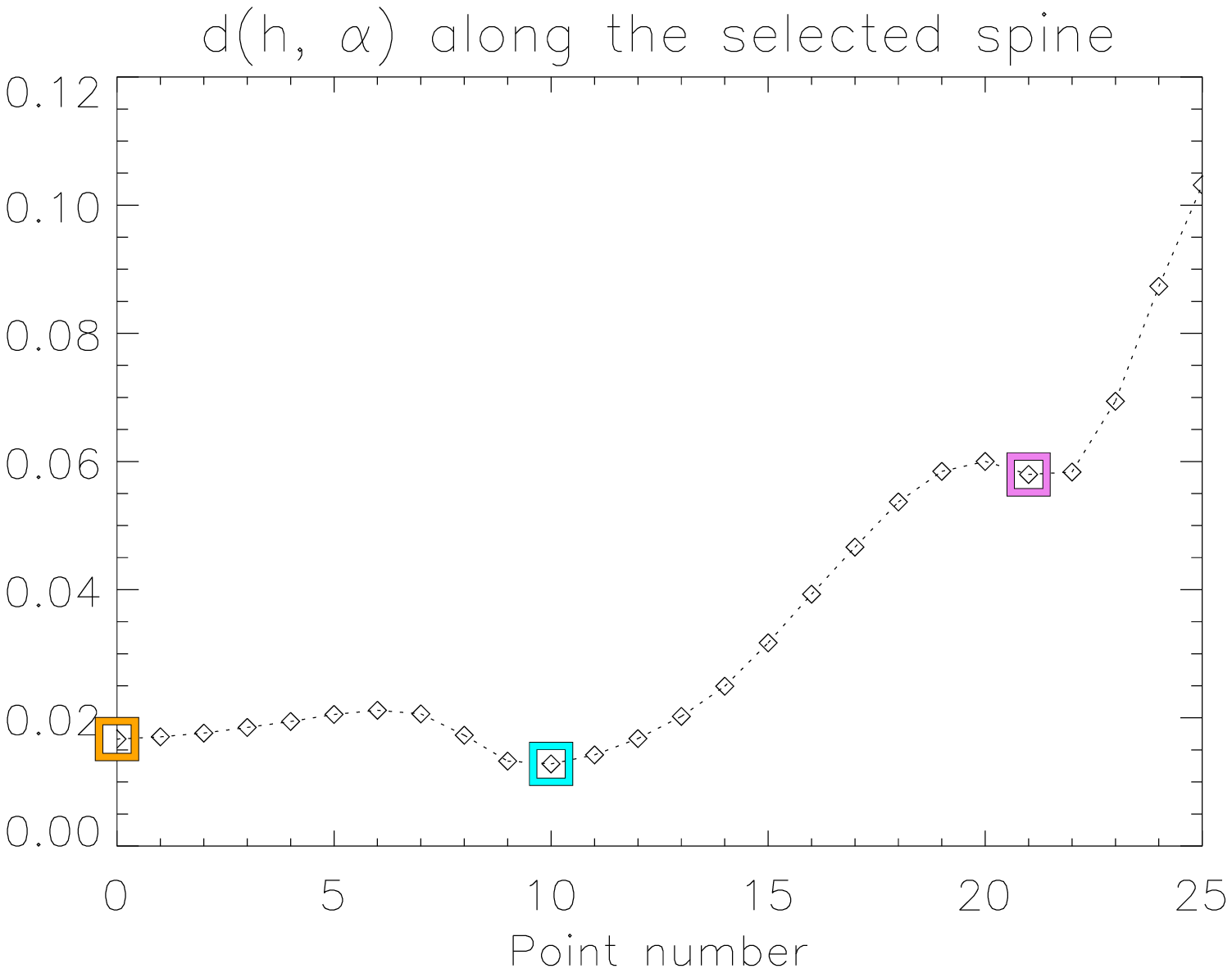} &
   (c)\includegraphics[width=9cm]{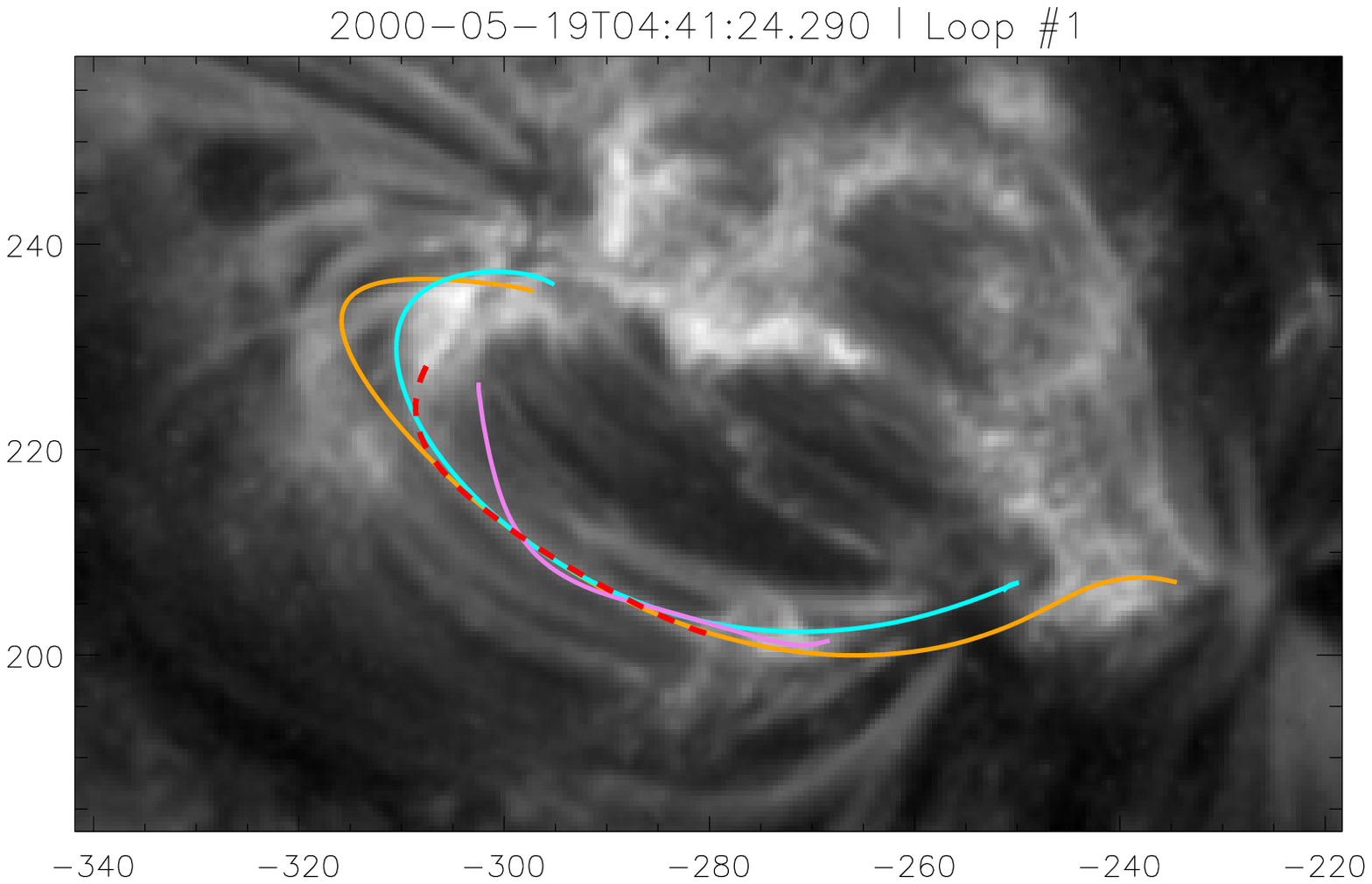} \\
  \end{tabular}
 \end{center}
 \caption{\small{A type of parameter space, not described in \citet{Malanushenko2009} on tests with analytic fields. (a) --- The parameter space. White diamonds: local minima in columns. White diamonds with red dots: selected ``non-hyperbolic'' valley. Three colored squares: points that correspond to the local minima on this valley. Blue dotted lines: hyperbolas $h=n\pi/\alpha$, $n=0, \pm1, \pm2, ...$. (b) --- $d(\alpha, h)$ along the selected valley. (c) --- A fragment of TRACE 171\AA~image with the traced loop (dashed red) and three field lines, corresponding to the three local minima. It seems that the one, corresponding to the lowest $h$ (magenta) is visually a much worse fit than the one, corresponding to the global minimum on this valley (cyan). We consider the valley to the left of the 'hump' a different valley, that could have been a degenerate hyperbolic. Based on this, we decide that the cyan line should be used. We update the fitting algorithm with the parameter space of this type.}}
 \label{new_minimum_type}
 \end{figure}

Among those ``underscribed'' types of parameter spaces there was one that was found to occur frequently enough ($\sim36\%$ of all loops) to warrant new classification. The ``anomalous'' (or ``non-hyperbolic'') valley in the parameter space of this type looked like two branches at each side of $\alpha=0$ line joined together (see Figure~\ref{new_minimum_type}). According to the above mentioned algorithm such a loop had to be either ignored or the local minimum on this valley that corresponded to lowest height was to be chosen. We found that such a choice results in a fit visually much worse than chosing the global minimum of this valley. We thus proceded with selecting the global minimum on the non-hyperbolic valley. 

This type of the parameter space had not been found on tests in analytic data and there is no clear indication that it is trustworthy, except for the better visual correspondence to the coronal loops. It might represent a distortion of ``conventional'' shapes due to the inclination of the line of sight. We would like to mention that such classification of parameter spaces is subjective and sometimes ambiguous, but it has found to yield statistically reasonable results when tested on analytical fields. As for the new type of solutions, we also would like to mention that many previous studies relied on visual comparisons. For example, \citet{Burnette2004} selected a linear force-free field with $\alpha_{best}$ that seemed to match visually many coronal loops; they repeated same procedure for many active regions and found that $\alpha_{best}$ is generally consistent with the average \als inferreded from vector magnetograms. In other studies that fit field lines to coronal loops by minimizing the average distance, such as \citet{Green2002, Lim2007}, the existence of different solutions was not given much attention. So while about $\sim$55\% of all our datapoints are backed up with tests on analytical fields, $\sim$36\% of them are obtained using methodology that has been previously used in many similar studies.
 
This procedure resulted in a ``best-fit'' line of a linear force-free field for every coronal loop. The quality of the fit was visually judged and assigned a subjective grade of C (poor fit), B (good) or A (perfect match). For further analysis, only loops that had a fit quality B or A were used. For example, the fit on Figure~\ref{new_minimum_type} was given a quality grade B and the fit on Figure~\ref{fig_3d} was judged to be of quality A.

\section{Results}
\label{sec_res}
We have performed the reconstruction procedure for the total of 303 coronal loops of AR 9004. 61 of them (20\%) had quality B and 219 (73\%) had quality A, so in total there were 280 (93\%) successful reconstructions. The real percentage of the successful fits might be slightly lower, as some coronal loops seeming to belong to AR 9004 on EUV image were reconstructed as open field lines or as field lines interconnecting ARs 9002 and 9004, and some of those fits have failed as well. We could have estimated how many loops seemed to have incorrect reconstructed connectivity, but we chose not to do so. The reason is that it would involve visual estimation of the connectivity that may or may not be right on its own. 

\begin{figure}[!hc]
 \begin{center}
  \begin{tabular}{p{4.5cm}p{4.5cm}p{4.5cm}p{4.5cm}}
   \includegraphics[width=5.0cm]{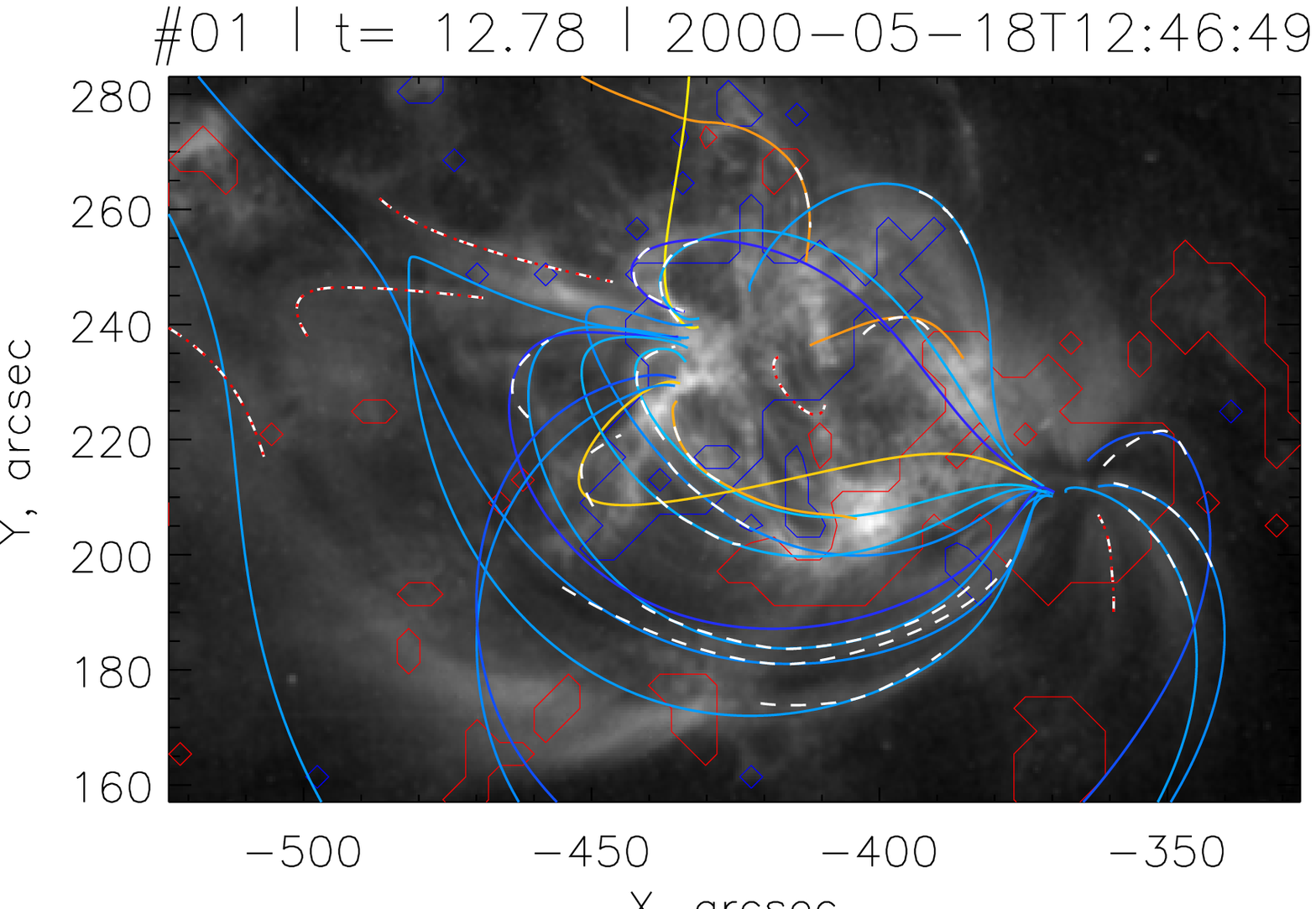} &
   \includegraphics[width=5.0cm]{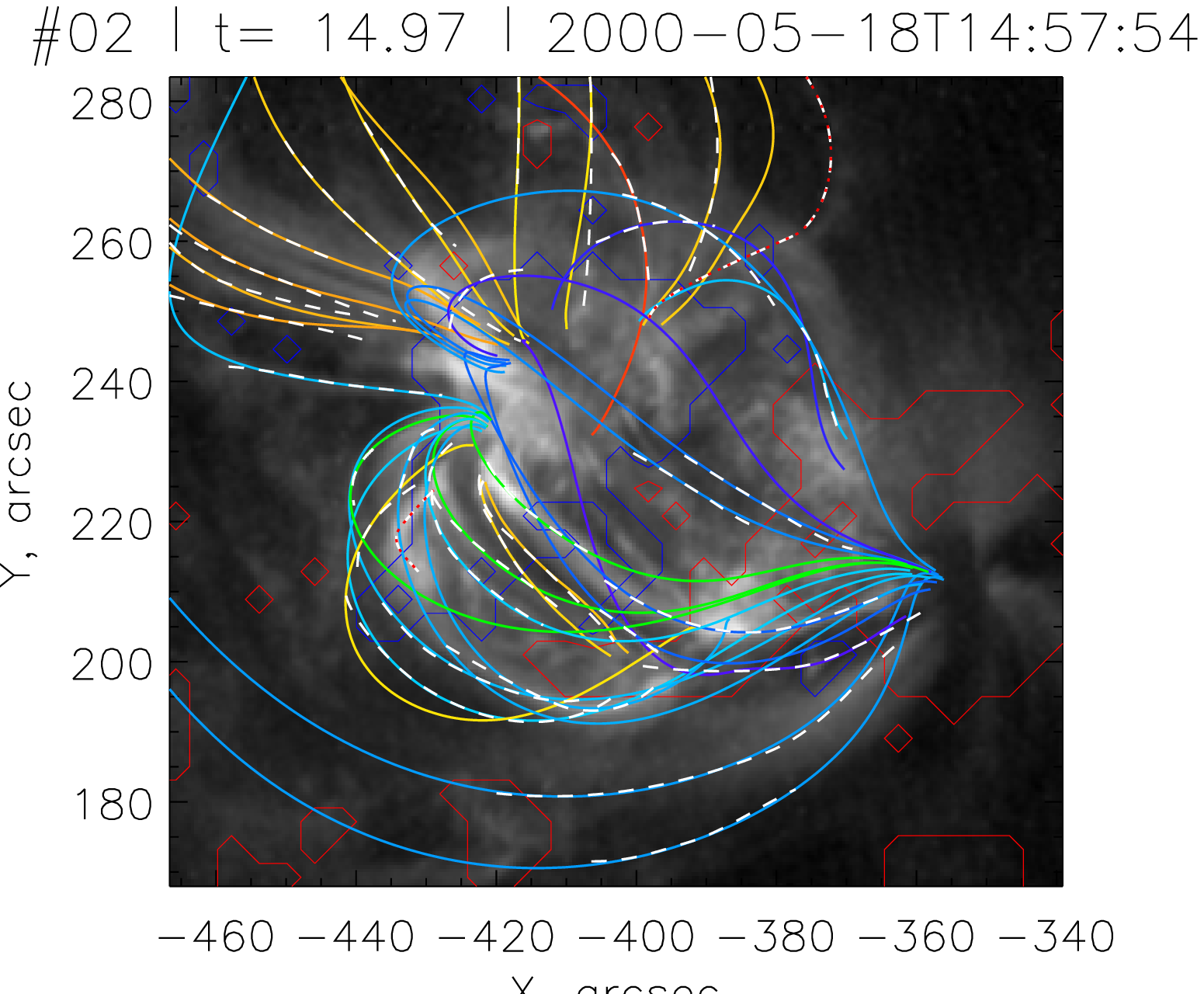} &
   \includegraphics[width=5.0cm]{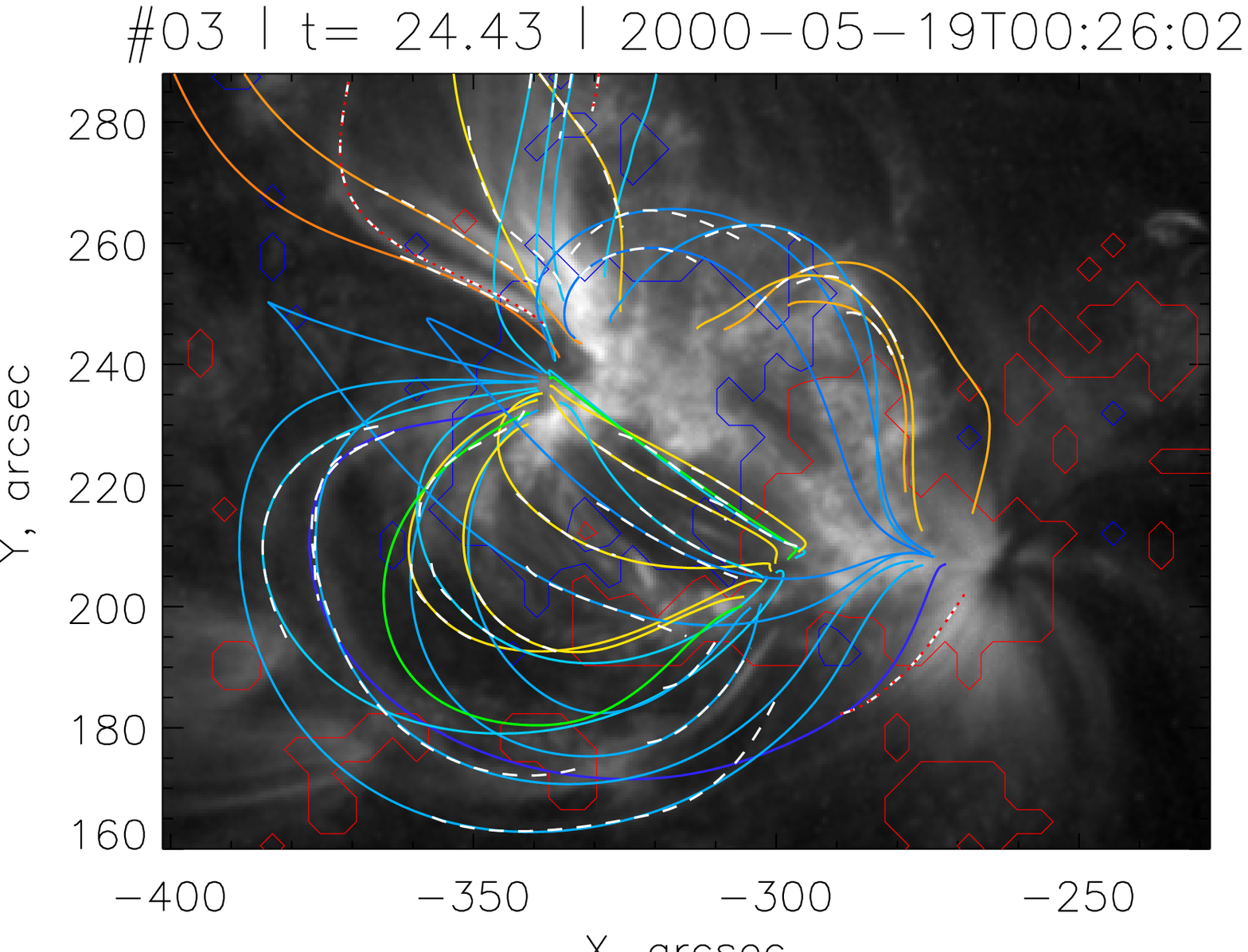} &
   \includegraphics[width=5.0cm]{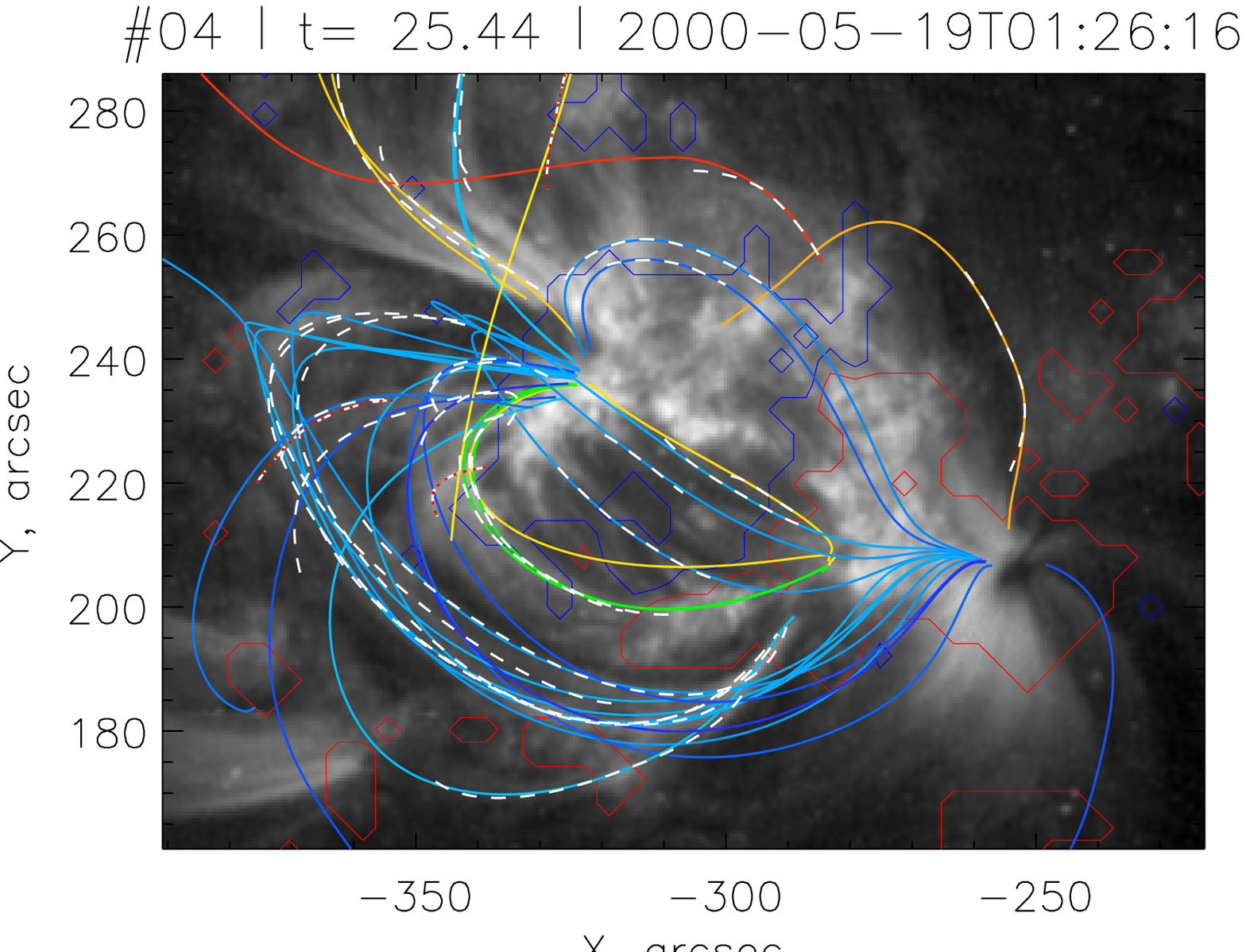} \\
   
   \includegraphics[width=5.0cm]{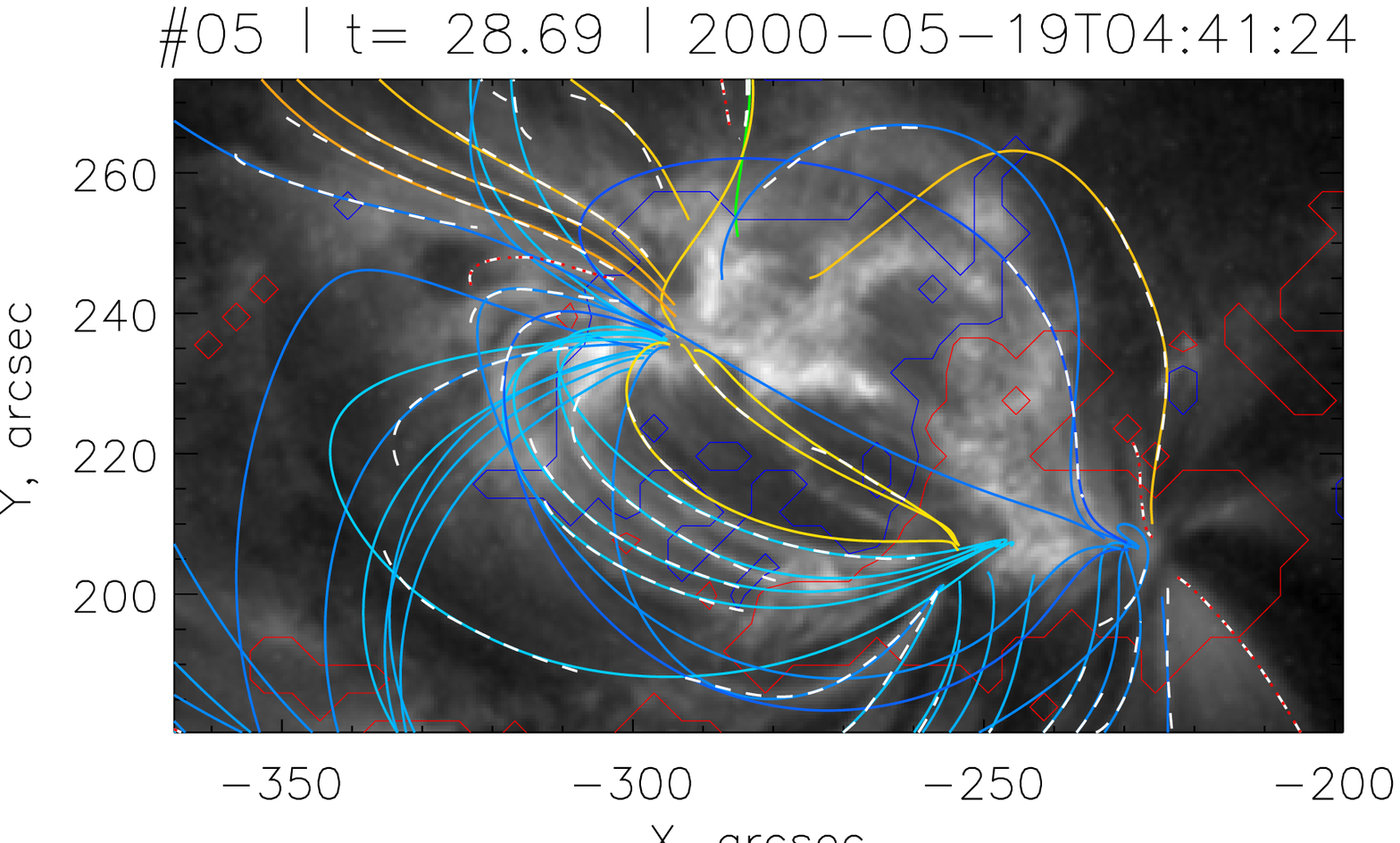} &
   \includegraphics[width=5.0cm]{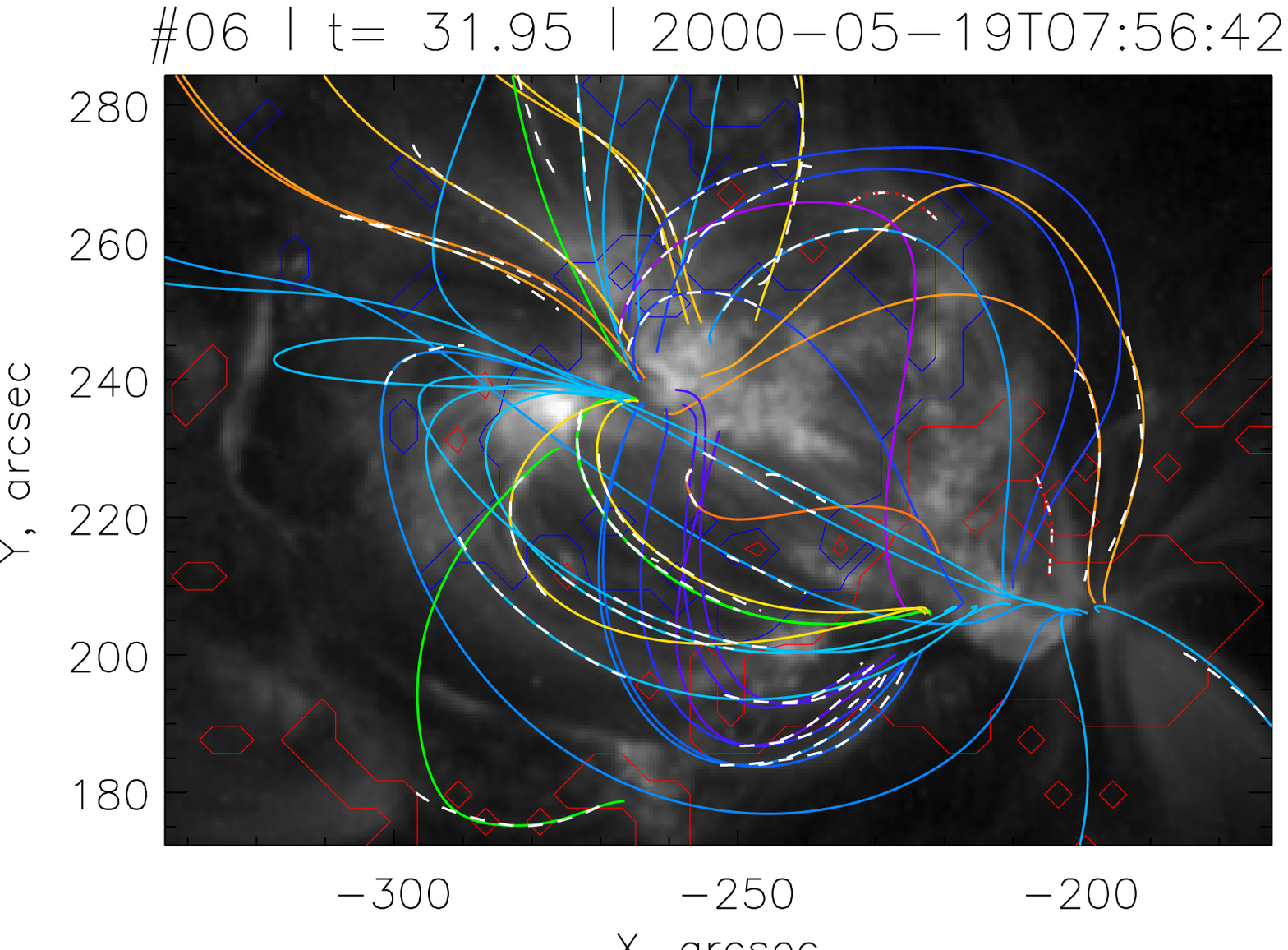} &
   \includegraphics[width=5.0cm]{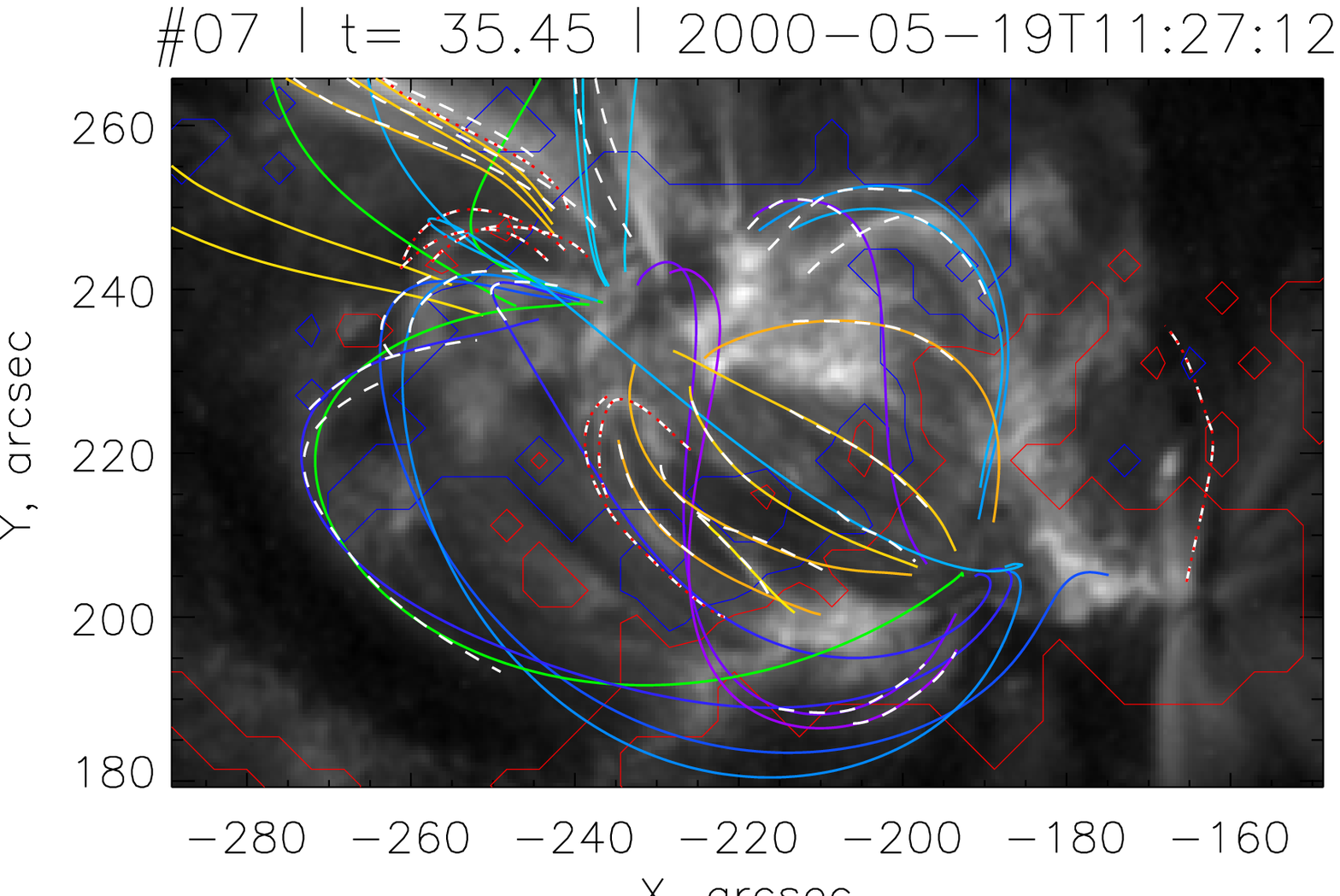} &
   \includegraphics[width=5.0cm]{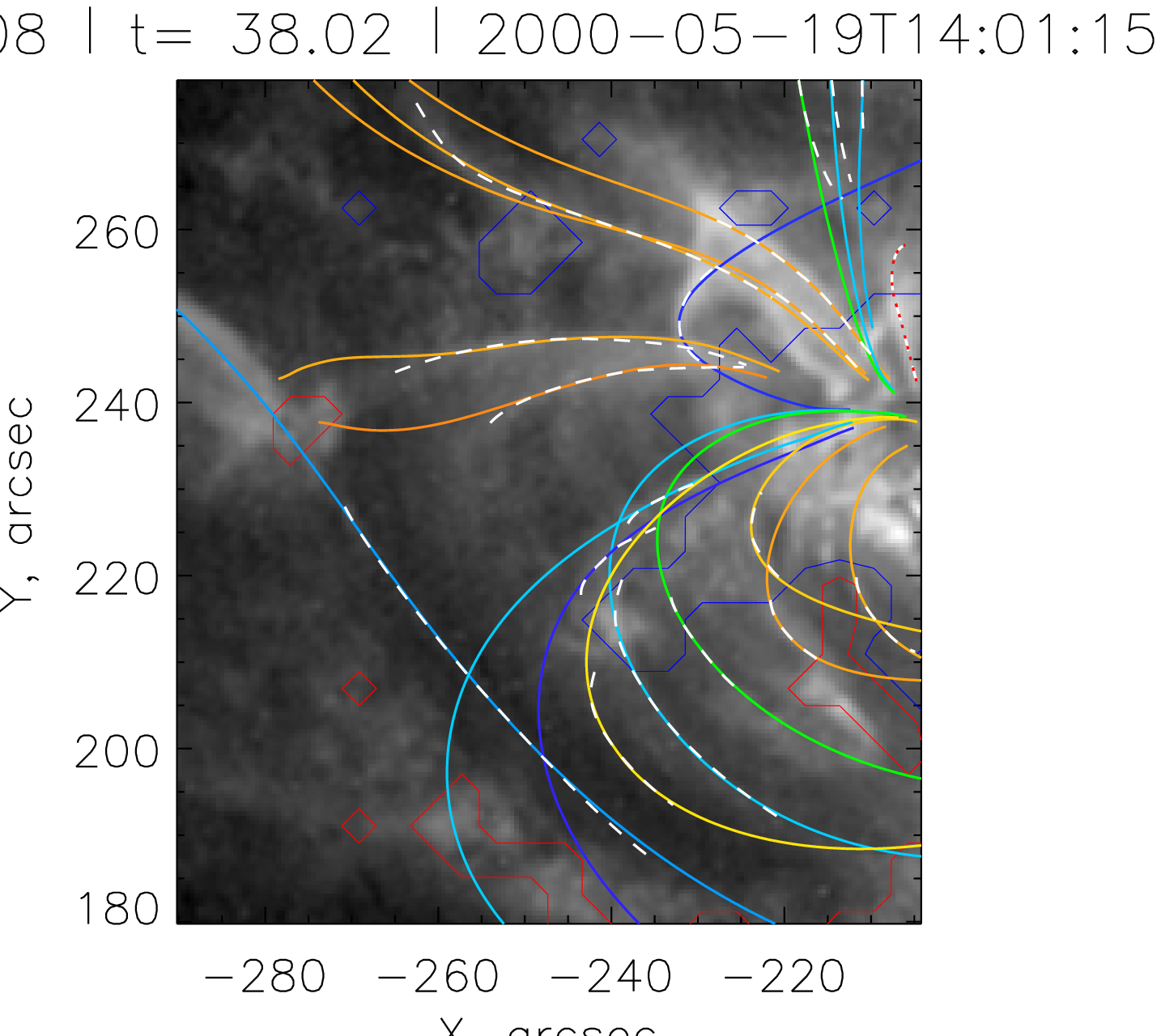} \\
   
   \includegraphics[width=5.0cm]{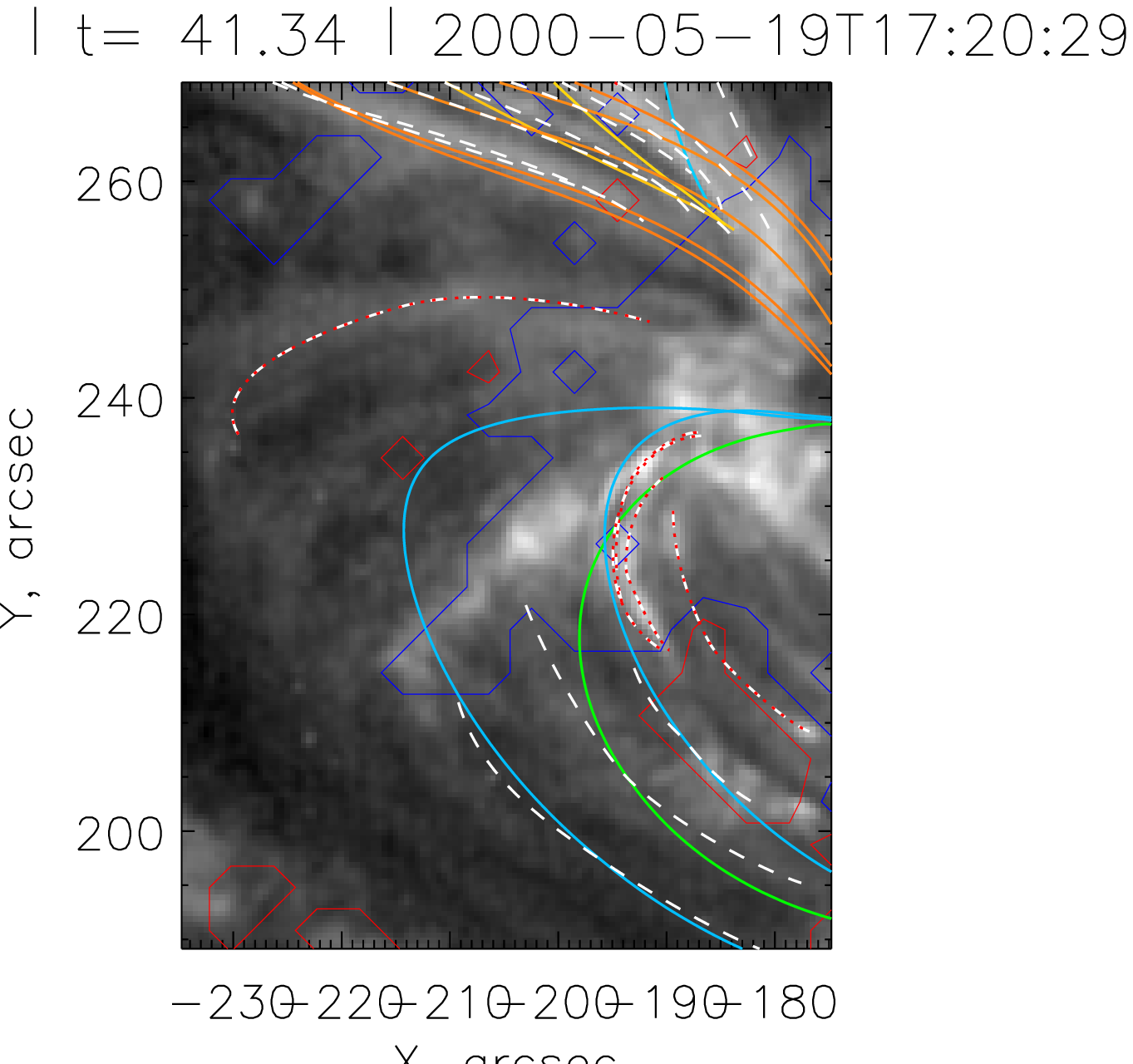} & 
   \includegraphics[width=5.0cm]{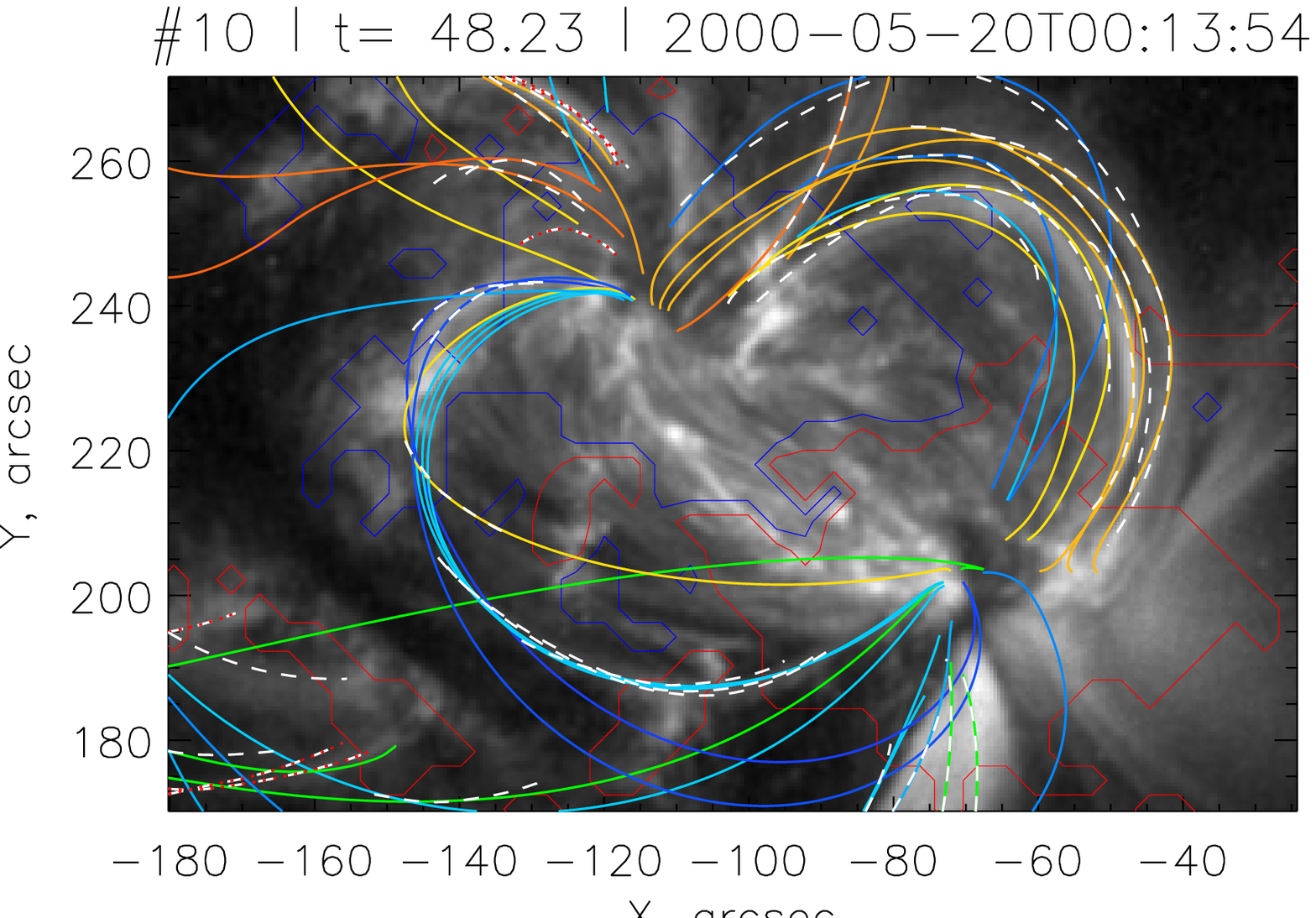} &
   \includegraphics[width=5.0cm]{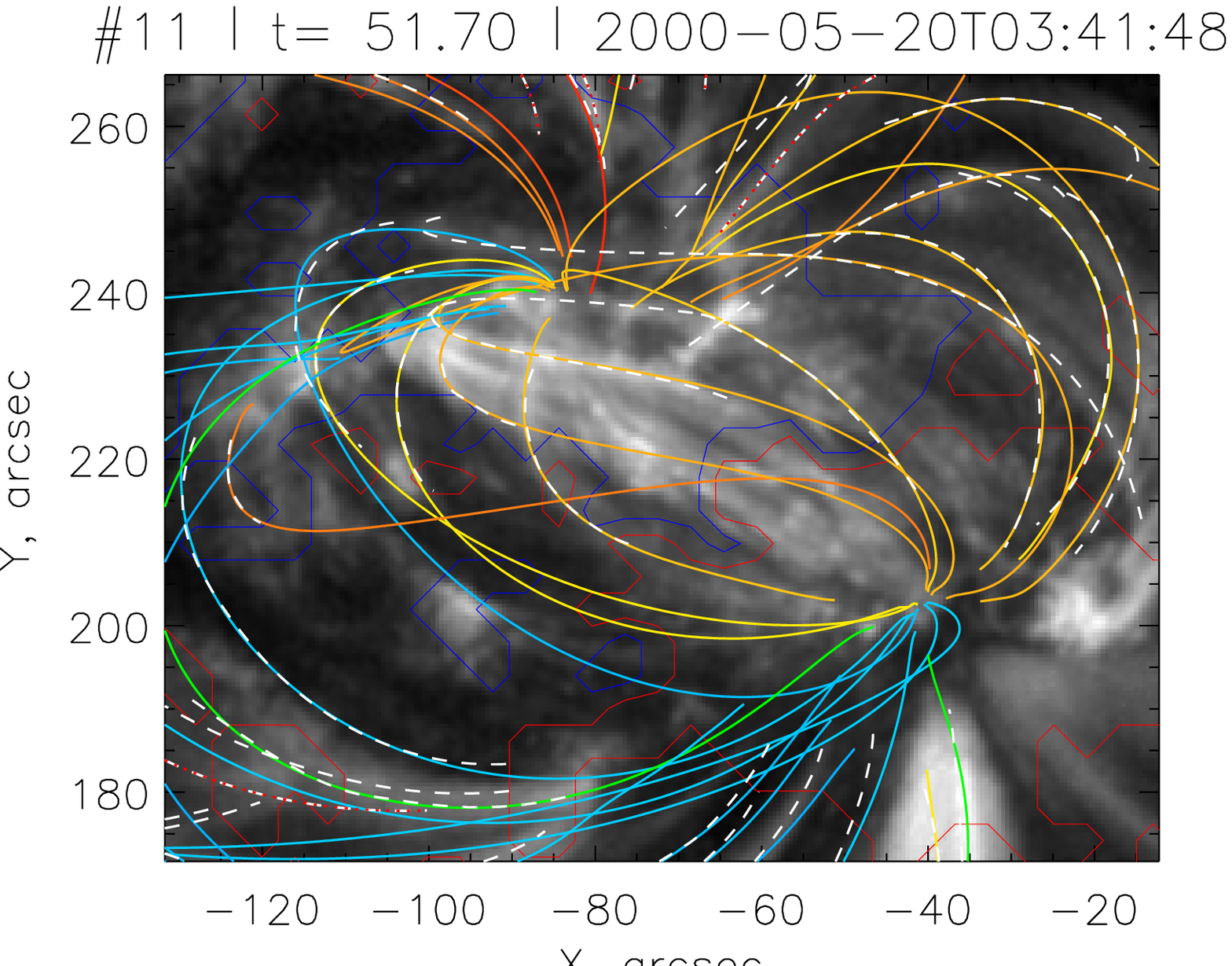} &
   \includegraphics[width=5.0cm]{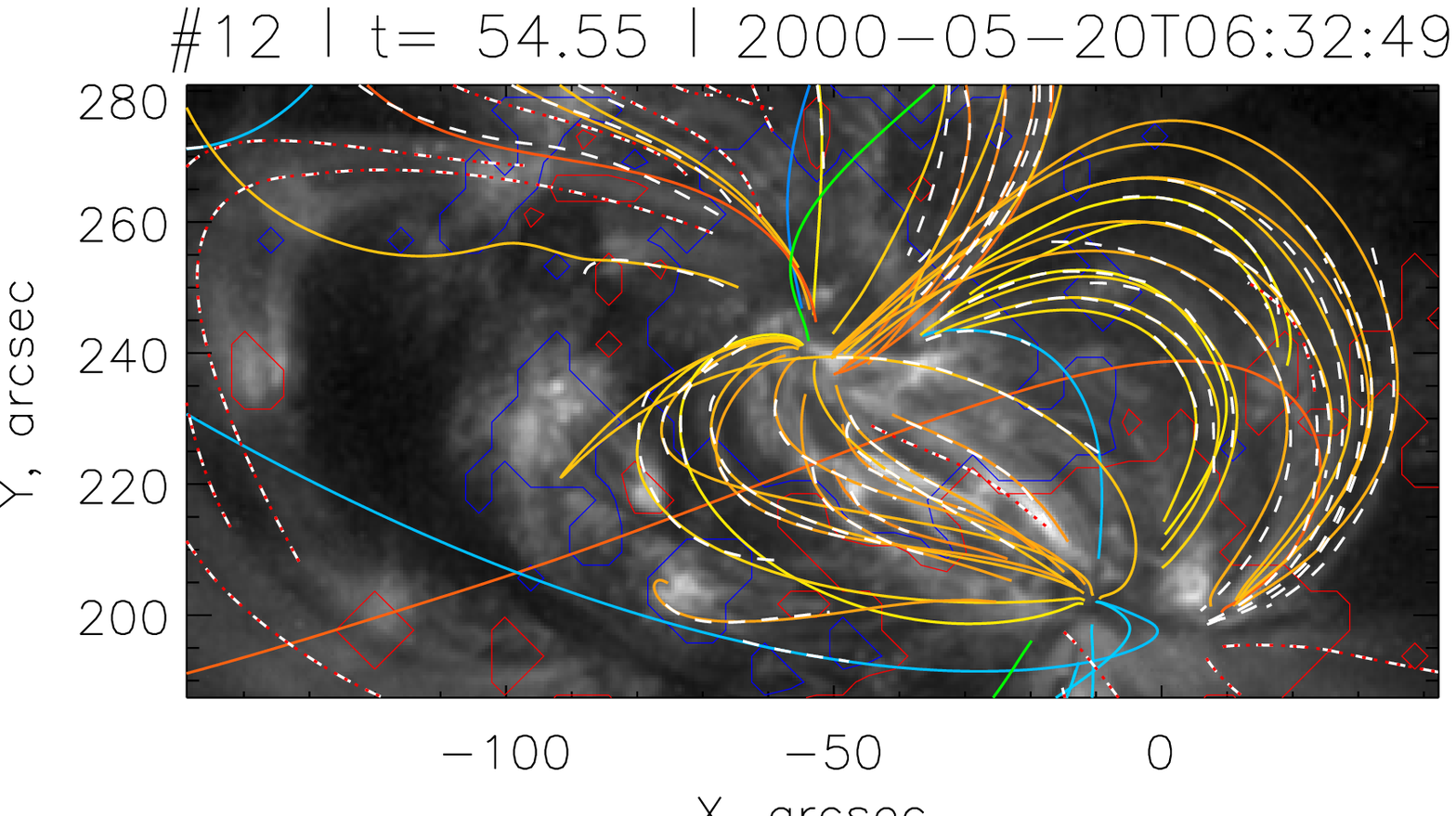} \\

   \includegraphics[width=5.0cm]{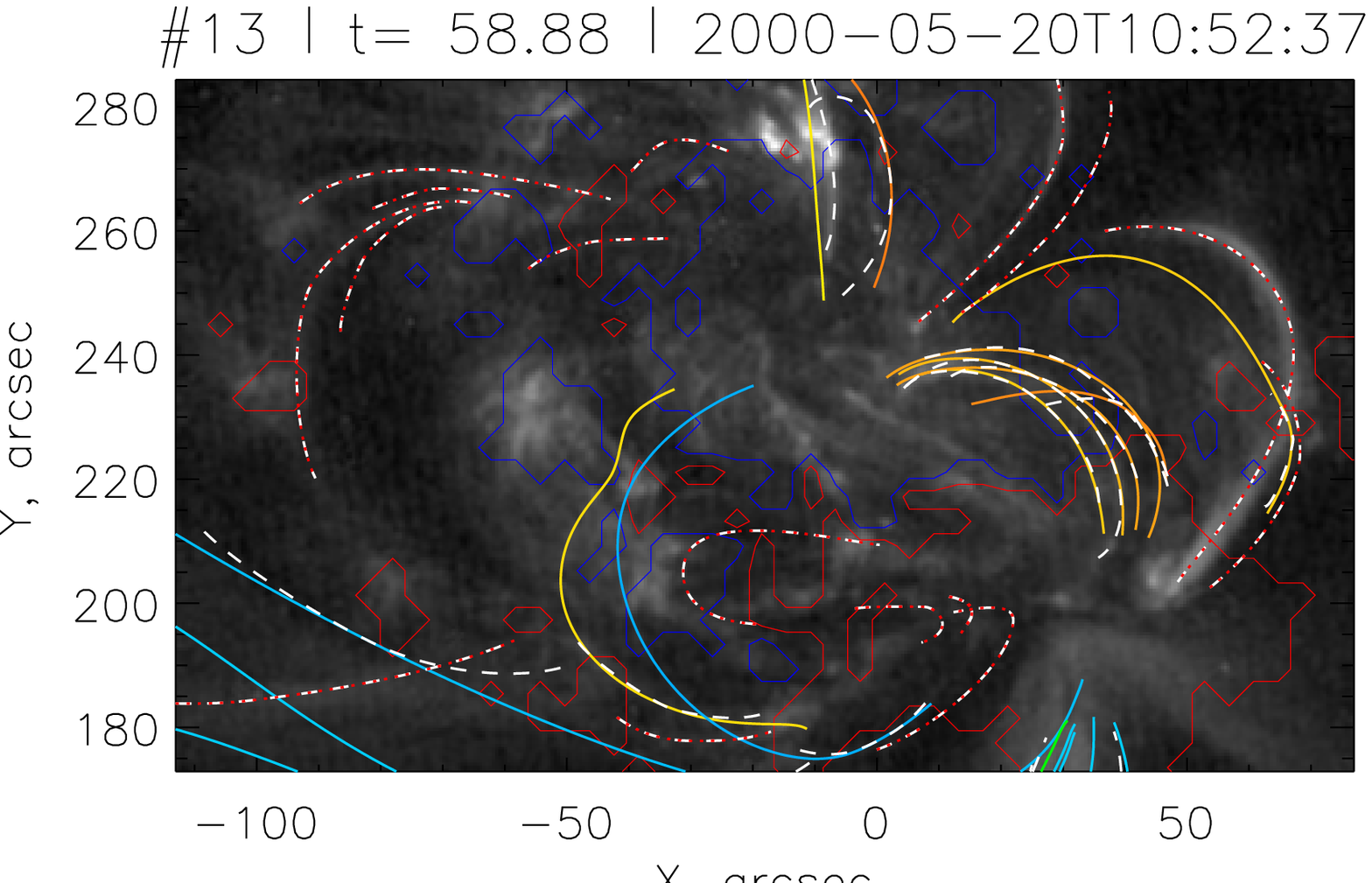} & 
   \includegraphics[width=5.0cm]{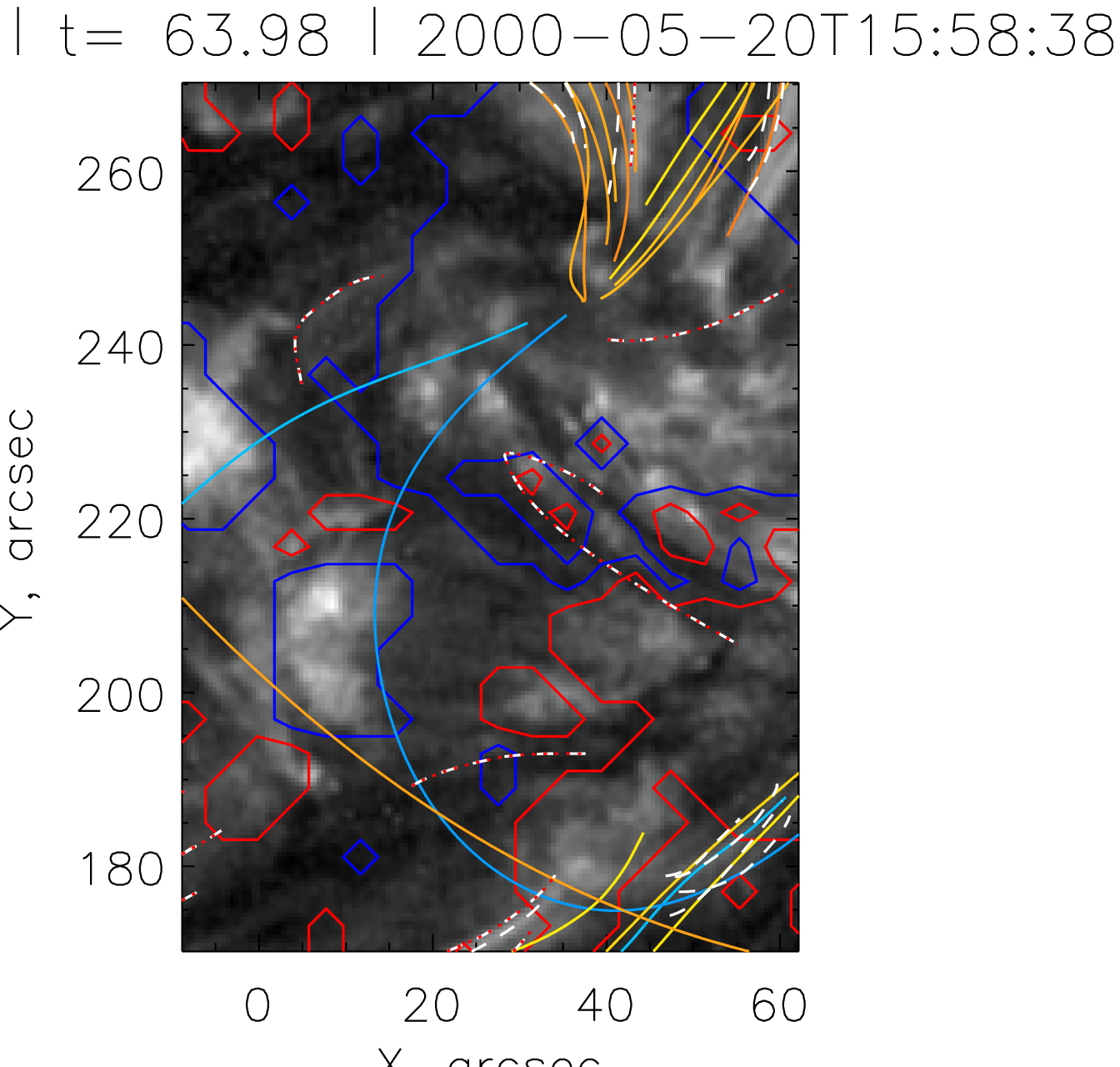} 
   \includegraphics[width=5.0cm]{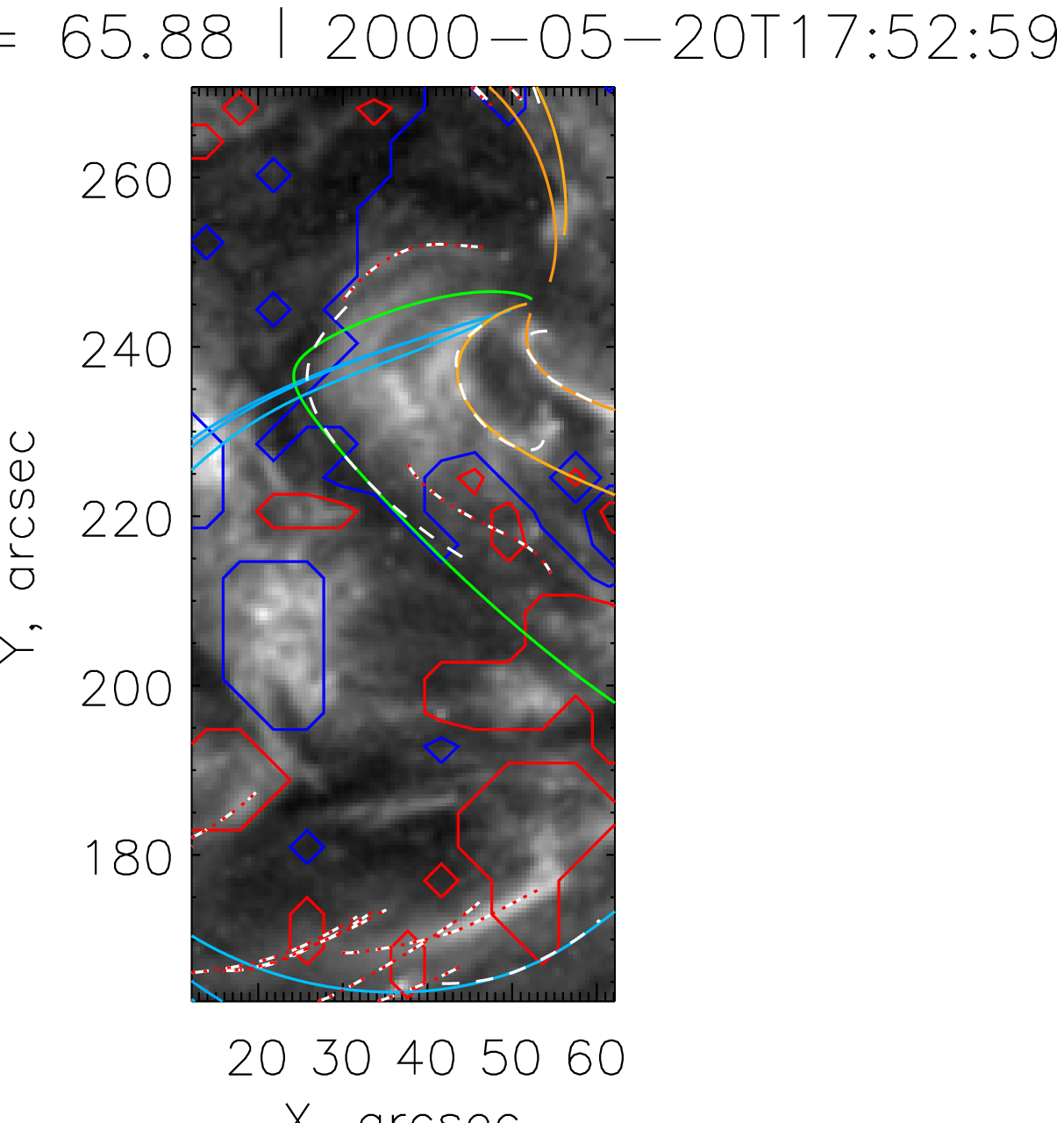} &    
   \includegraphics[width=5.0cm]{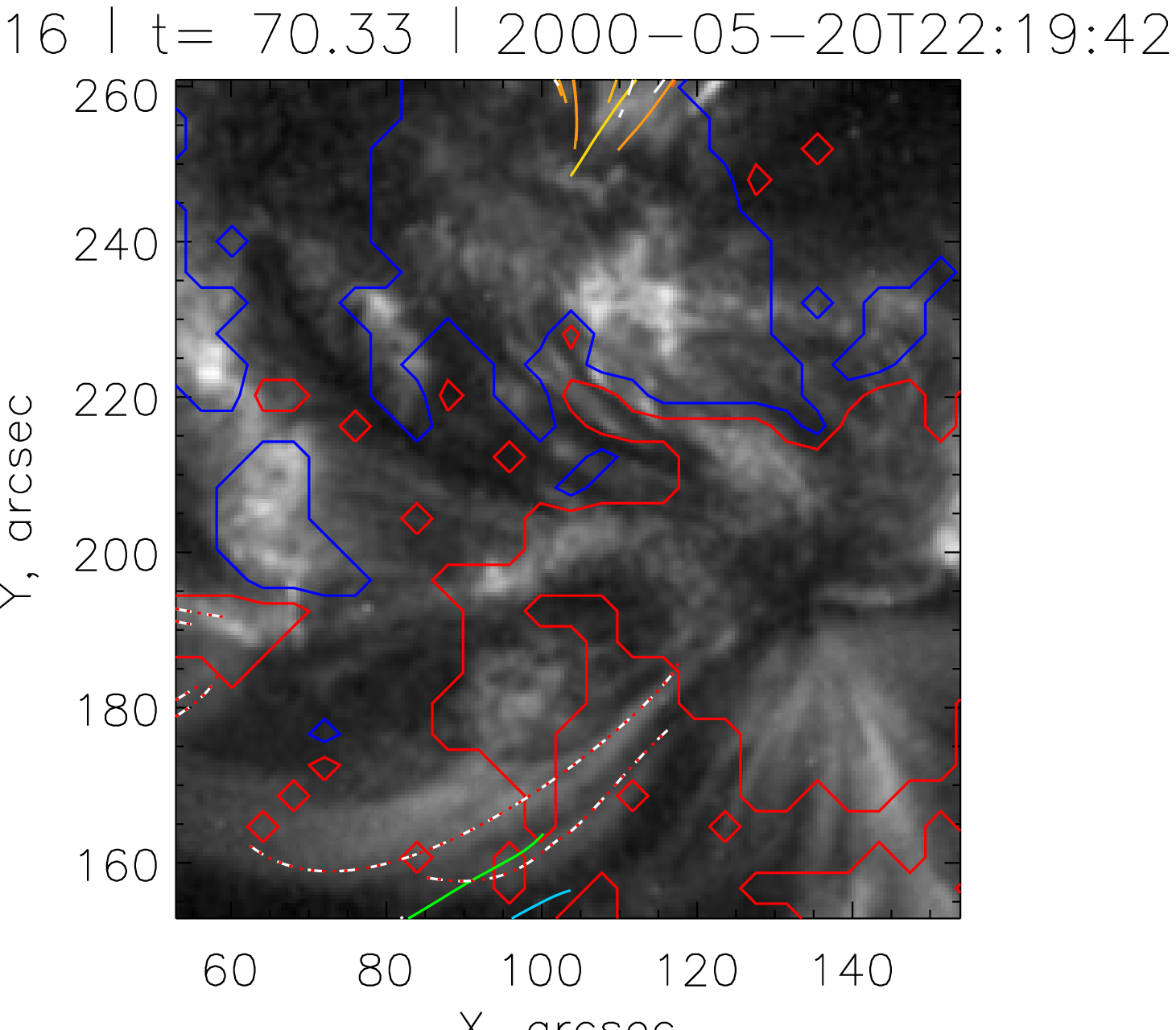} & 
   \includegraphics[width=5.0cm]{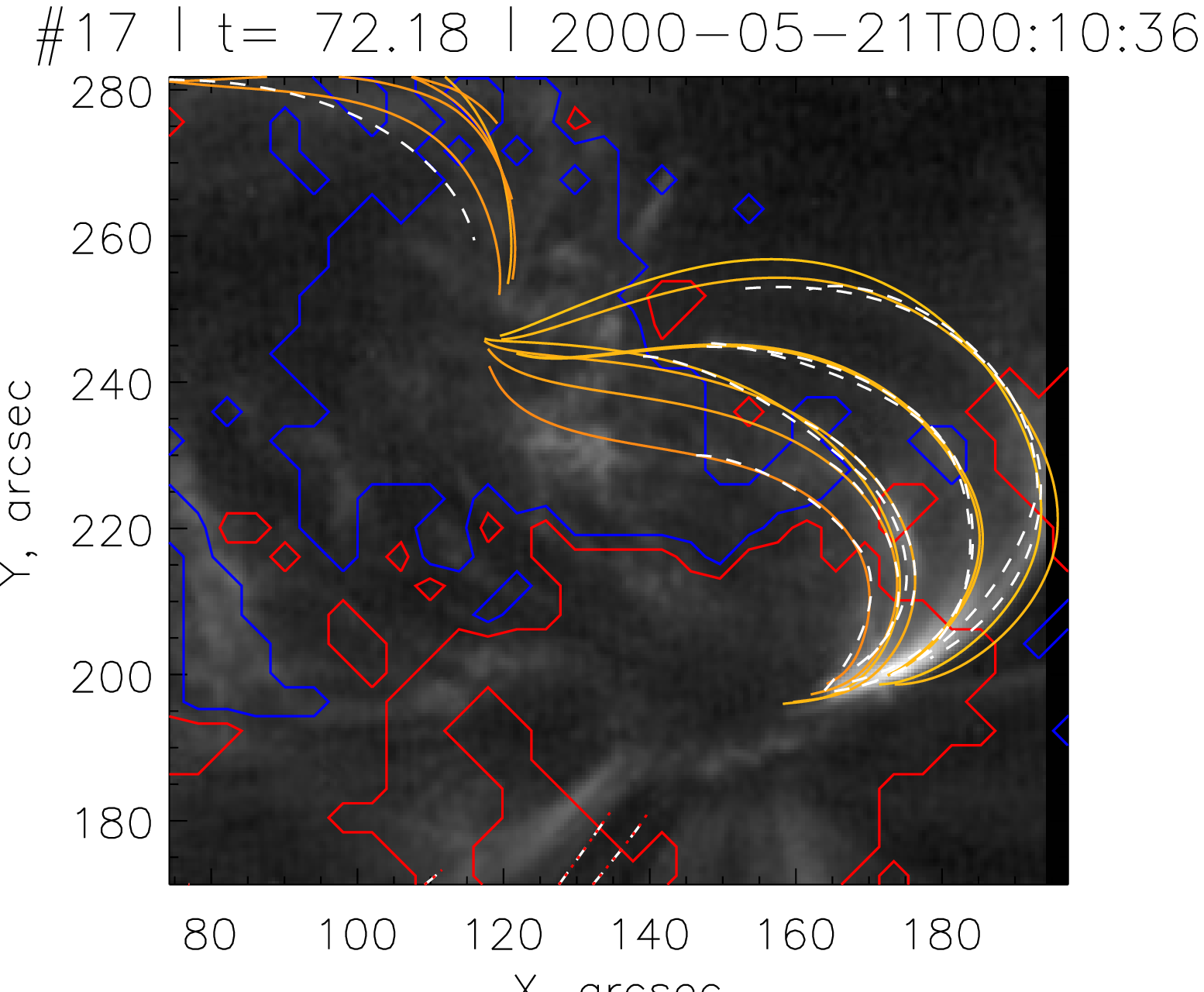} \\
   
   \includegraphics[width=5.0cm]{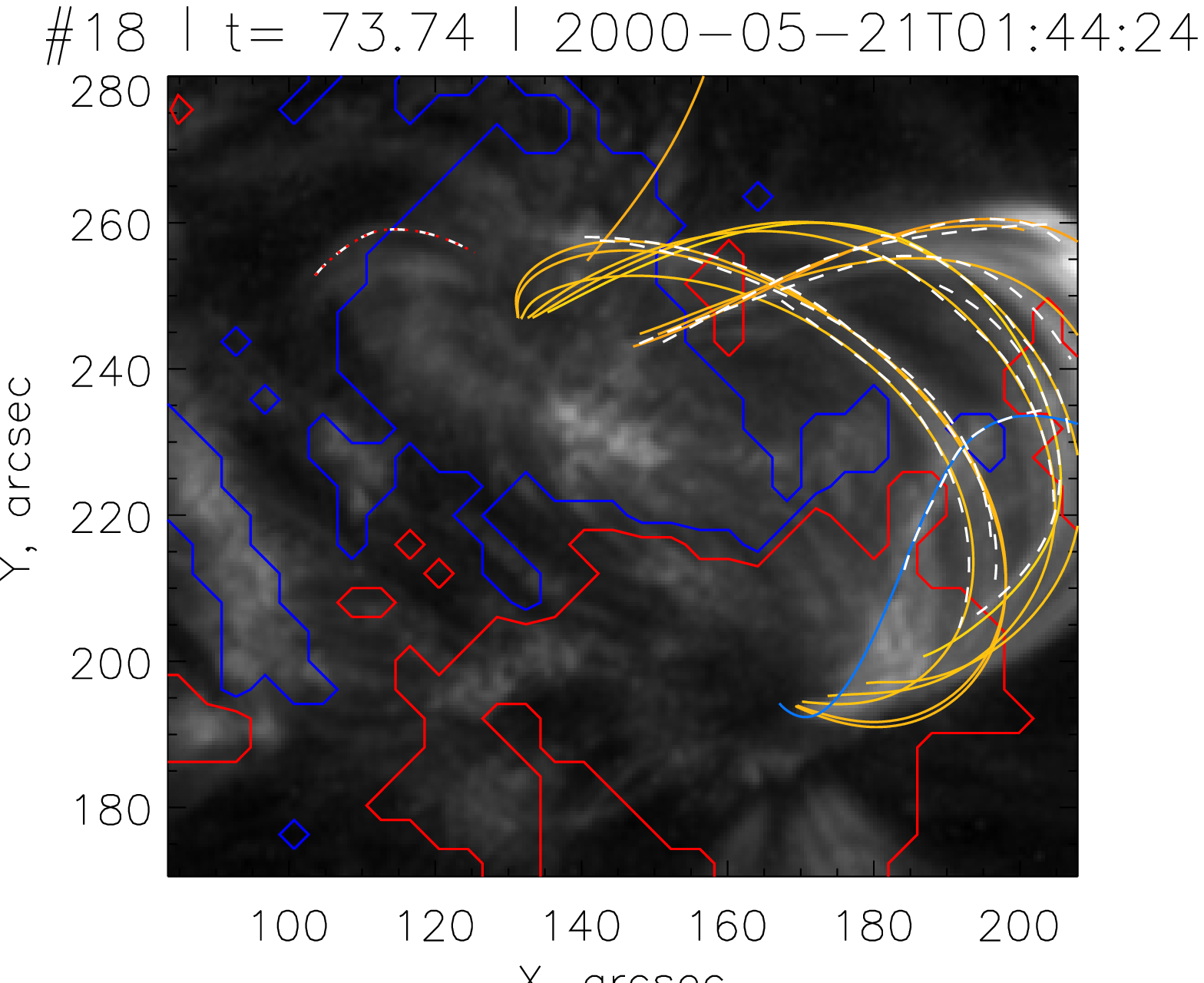} &
   ~\hspace{.5cm}\includegraphics[width=5.0cm]{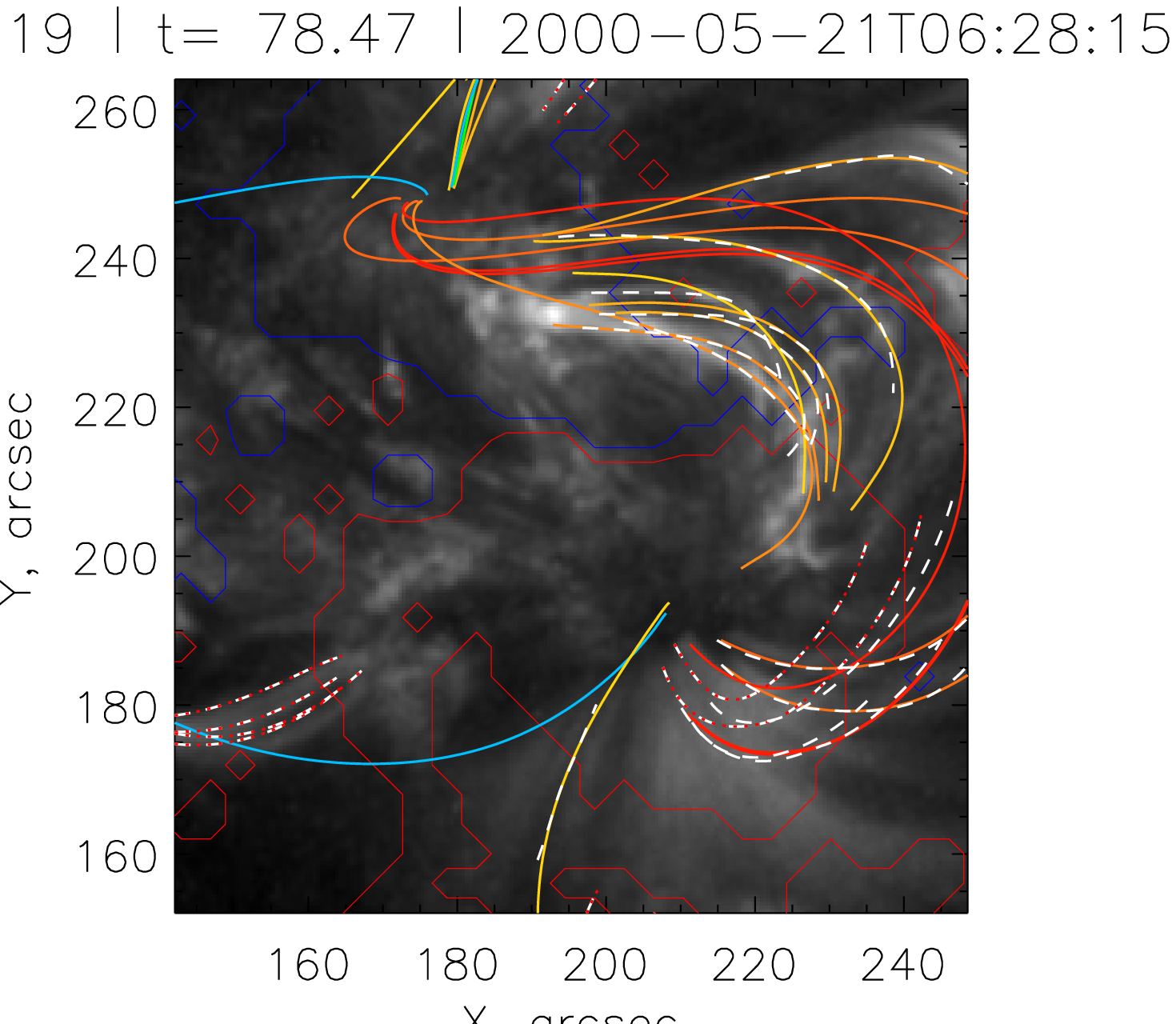} &
   \includegraphics[width=5.0cm]{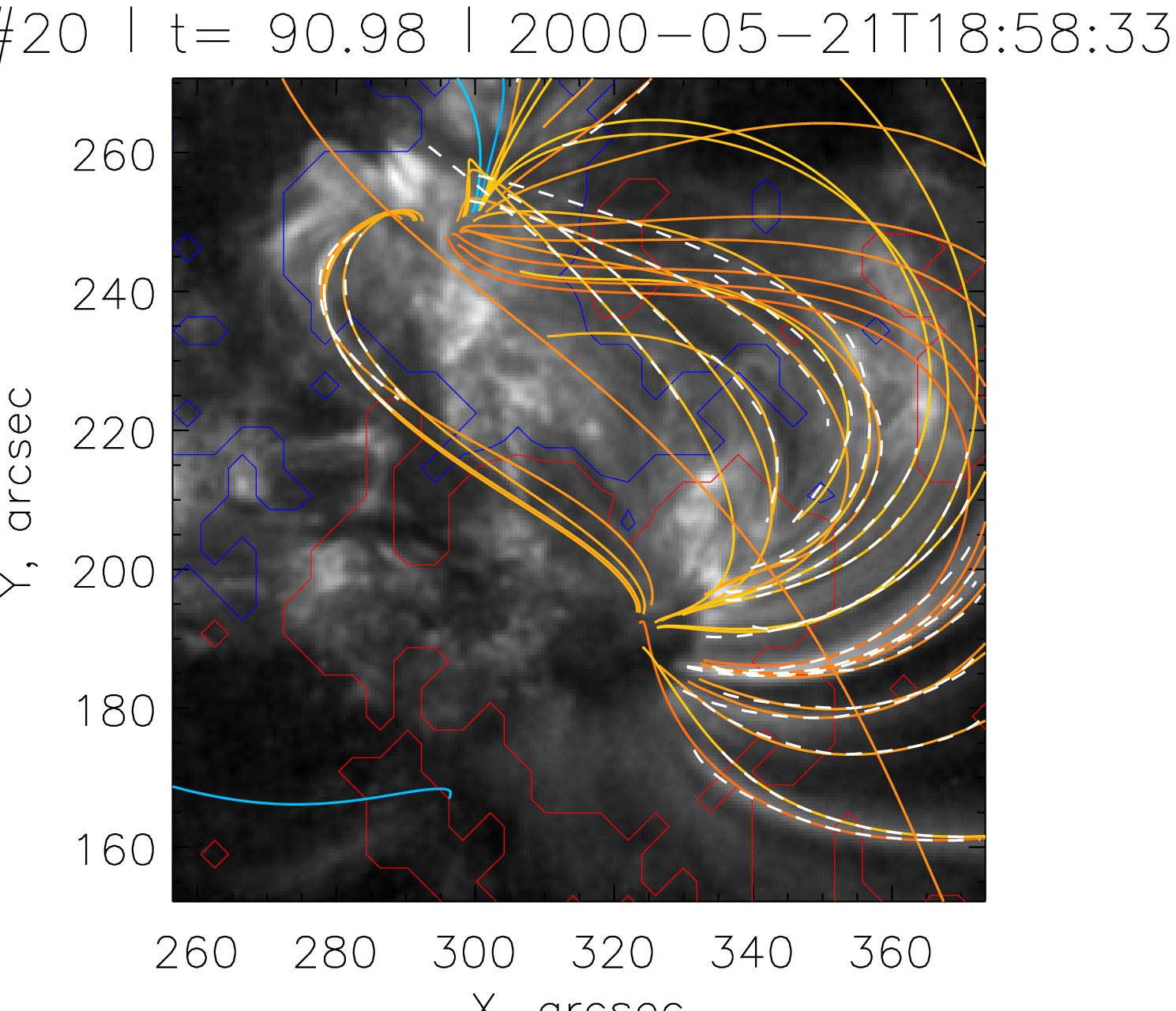} &
    \includegraphics[width=5.0cm]{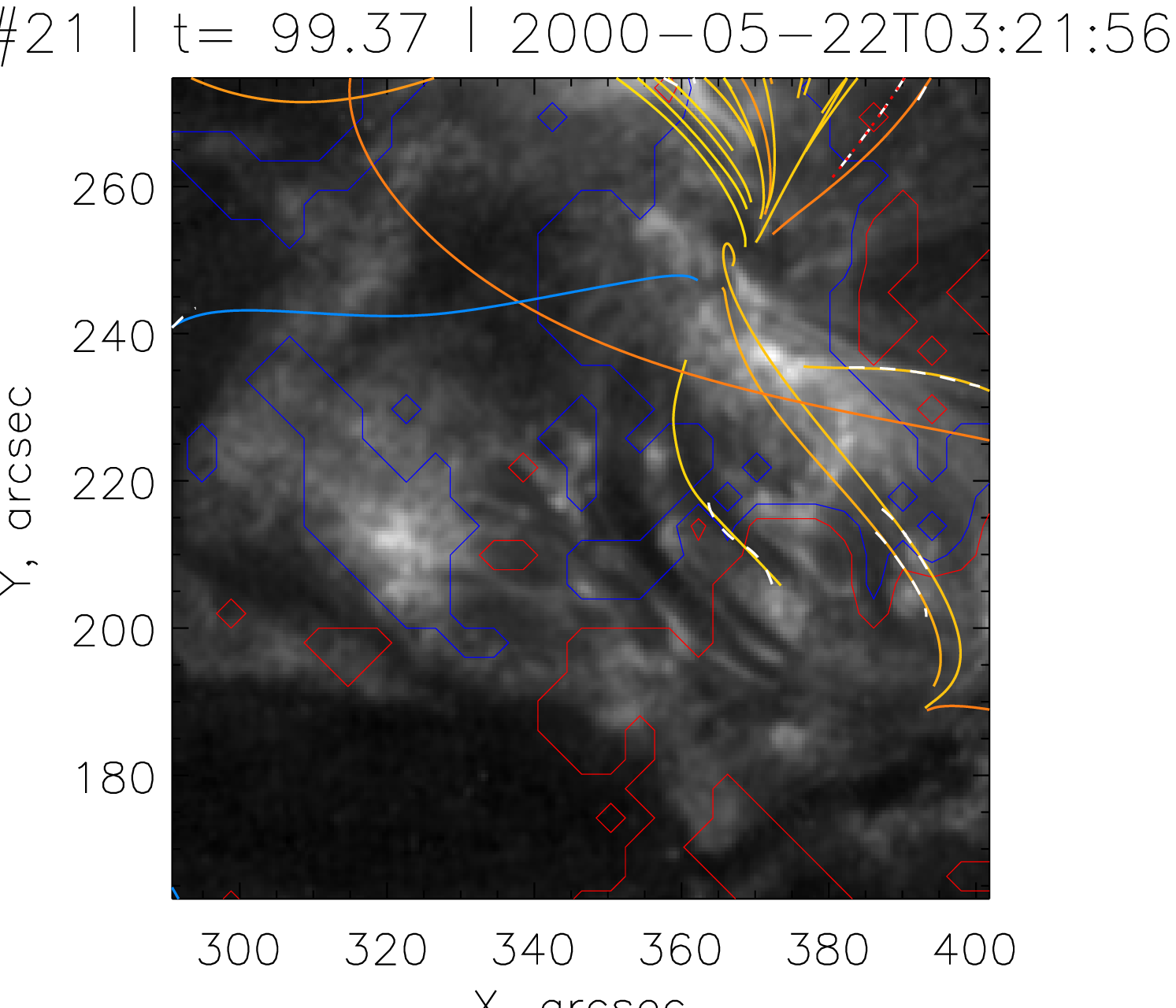} \\
   &&\\
   \multicolumn{3}{c}{\includegraphics[width=7cm]{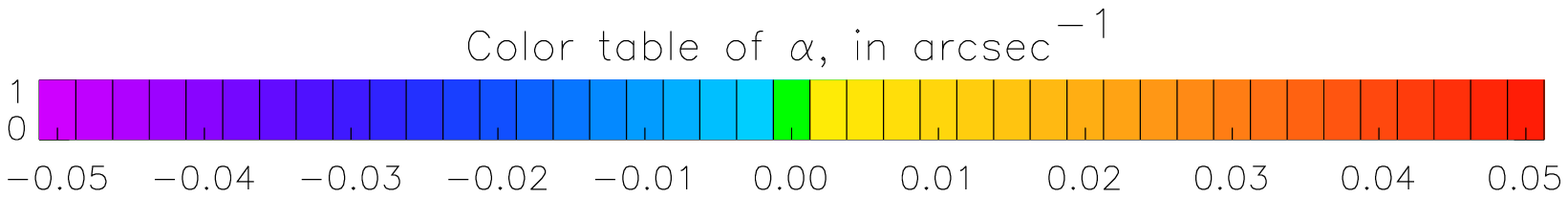}}\\
  \end{tabular}
 \end{center}
 \caption{\small{All reconstructed field lines of quality B or A (solid lines). Their color corresponds to \al. The original coronal loops are shown as dashed white (for successful reconstruction) or dashed red-white (for unsuccessful).}}
 \label{all_xrt}
 \end{figure}

The following evolution was observed: 
\begin{itemize}
	\item{Up to $t=35.5$ hrs. All times are in hours since 2000-05-18 00:00:00. Images 1-7 on Fig.~\ref{all_xrt} the loops had \textit{mainly negative (left-handed) twist} and the median twist has decreased in magnitude.}
	\item{From $t=48.2$ hr to $t=51.7$ hr (images 10-11 on Fig.~\ref{all_xrt}) had \textit{twist of both signs}, with negative twist predominantly in the eastern half of the dipole and positive twist predominantly in the western half. The median was close to zero. (Only half of the region was seen on images 8 and 9, so we could not draw any conclusion about spatial structure of the twist from $t=38.0$ hr to $t=41.3$ hr.)} 
	\item{Most of the loops on image 12, $t=54.6$ hr, had twist of positive sign.}
	\item{The loops on image 13, $t=58.9$ hr, were \textit{poorly fit}. As suggested in \citet{Malanushenko2009}, this might indicate that the field was strongly twisted or maybe strongly non-linear.} 
	\item{Almost no loops of AR 9004 were observed and well fit on images 14-16 ($t=64.0$ hr to $t=70.3$ hr). It is possible that the existing loops were \textit{outside} of TRACE field of view.} 
	\item{Most of loops on images 17-21 ($t=72.2$ hr to $t=99.4$ hr) had twist of positive sign (or right-handed twist). The median twist appeared to be slightly increasing.} 
\end{itemize}

\begin{figure}[!hc]
 \begin{center}
   \includegraphics[height=7cm]{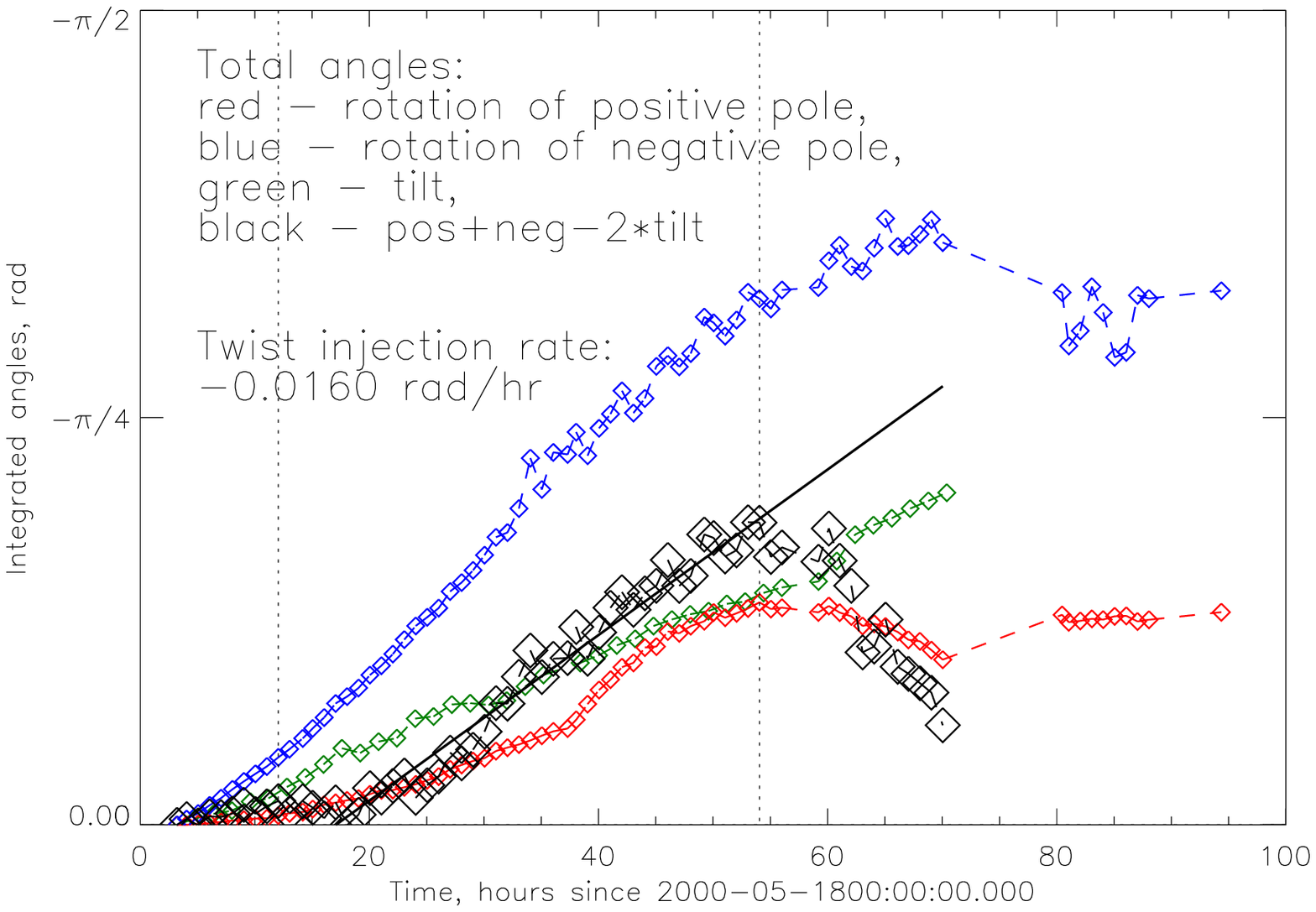}
 \end{center}
 \caption{\small{Twist rate, as measured by \citet{Longcope2007b}. The data points within time interval from $t_0=12$ hours to $t_1=57$ hours were fit to a line with least absolute deviations. The beginning of the time interval was chosen to be the lower boundary for (n)lfff data and the end of the time interval was chosen where the linear twist injection was no longer obvious.}}
 \label{longcope_twist}
 \end{figure}

The emergence of AR 9004 was evident from the evolution of the magnetic flux. The magnetic flux, according to \citet{Longcope2007b}, was steadily increasing until about $t=55$ hrs, and after that it exhibited a slight decrease. This decrease was concurrent with the drop in the combined twist angle on Fig.~\ref{longcope_twist}. 

For any given time there was a wide distribution of measured twist. There was, however, a general trend for the twist to increase. While most of the loops were negatively twisted at early times, most of loops at later time exhibit positive twist. 

\begin{figure}[!hc]
 \begin{center}
  \begin{tabular}{p{8.0cm}p{8.0cm}}
   \includegraphics[width=7.cm]{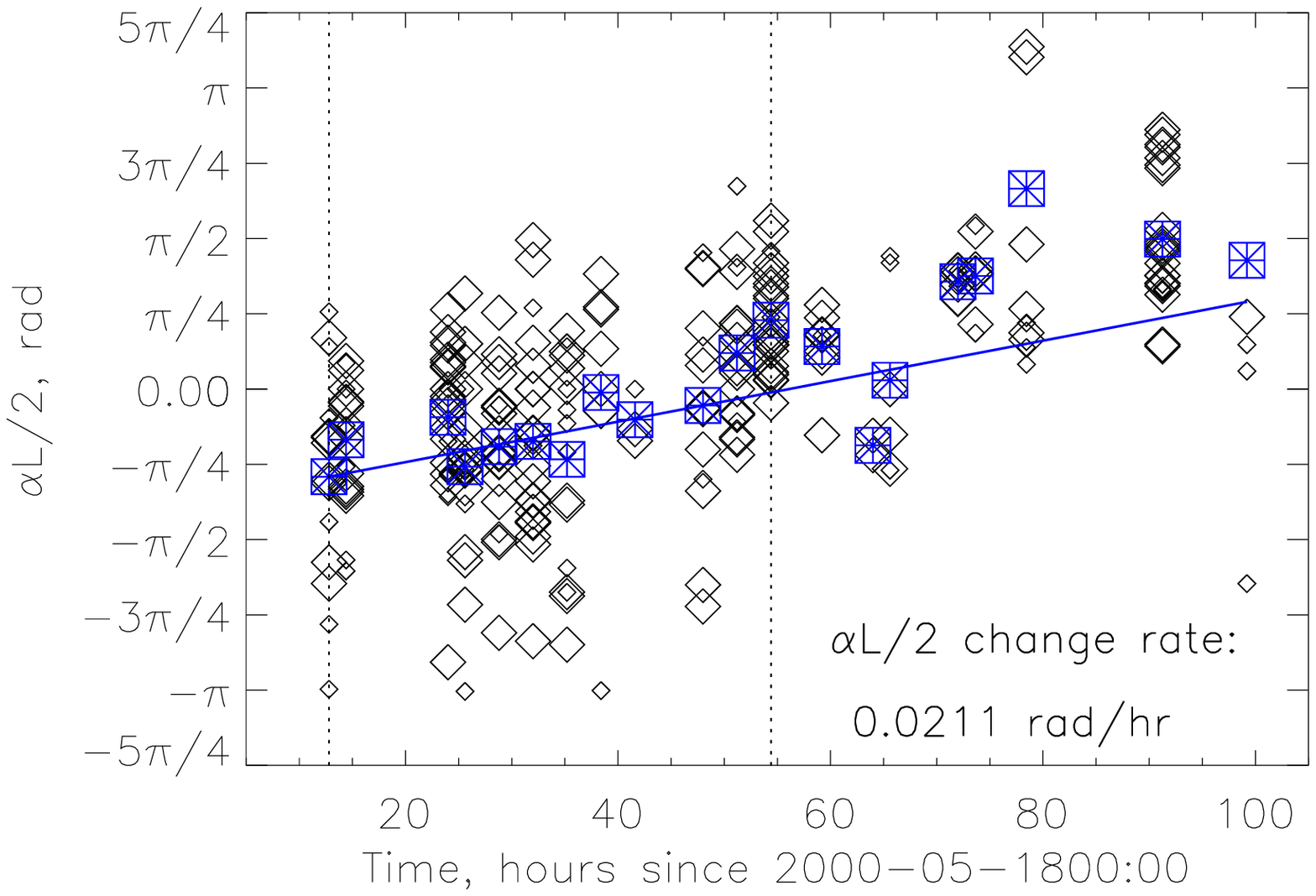} &
   \includegraphics[width=7.cm]{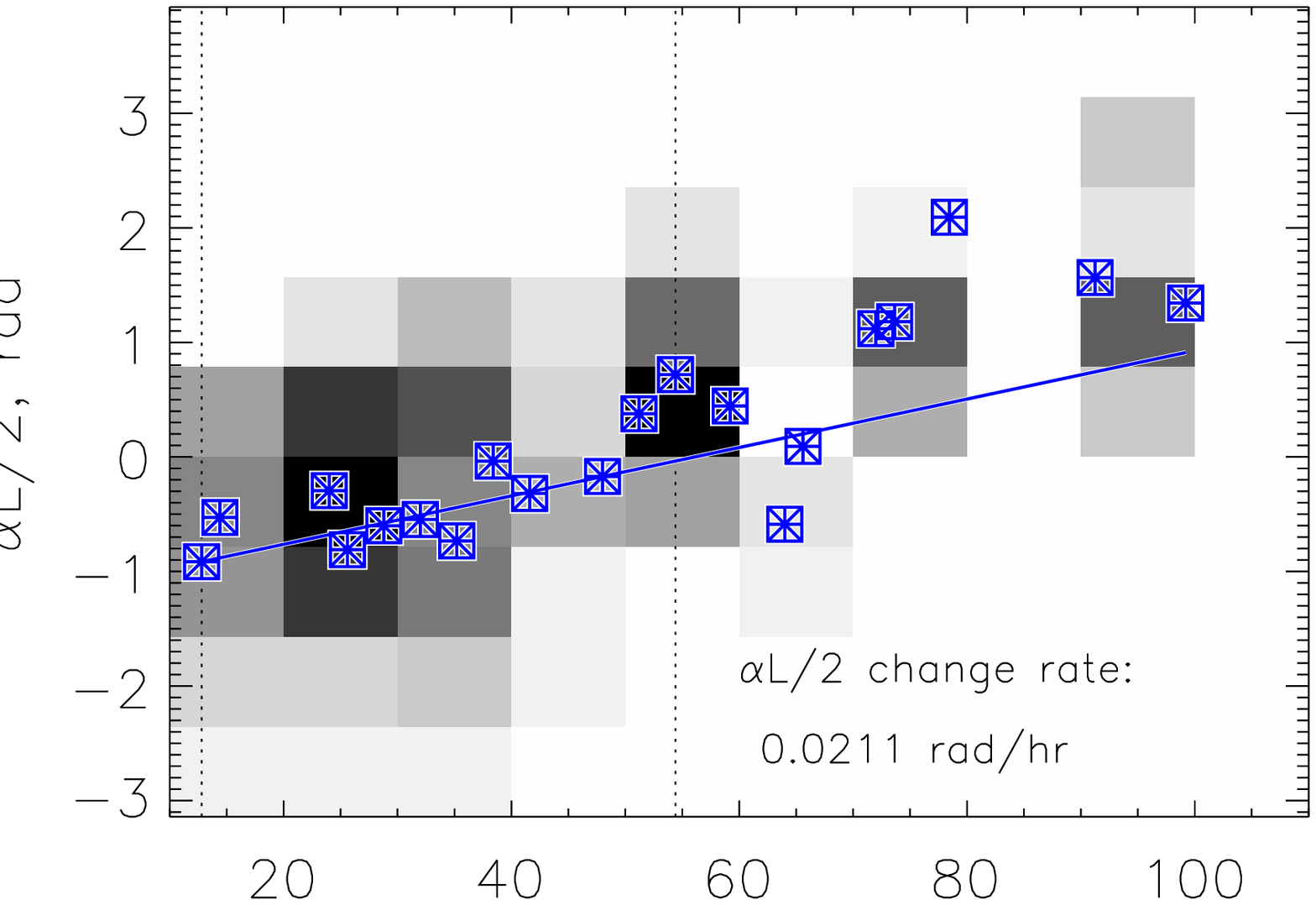} \\
   &\\
  \end{tabular}
 \end{center}
 \caption{\small{Twist of coronal loops versus time, using (n)lfff reconstruction from \citet{Malanushenko2009}. \textit{(left)} --- Diamonds show twist of individual loops (larger correspond to quality A and smaller to quality B). Blue squares are medians for each individual time. The line shows least absolute deviation to the means with means of the absolute deviations fit to the diamonds within the selected time range (same range as on Figure~\ref{longcope_twist}). \textit{(right)} --- A histogram showing time-twist distribution of coronal loops makes the trend evident. Black color corresponds to 20 points or more and white corresponds to one or no points.}}
 \label{alphaL}
 \end{figure}

We summarize the findings in Figure~\ref{alphaL}. Each point there represents $\alpha L/2$ of a single loop versus time when it was measured. At every time we find a mean twist and mean of the absolute deviation (this way the time frames with many data points would not be given more weight in the fitting procedure than the time frames with few data points). We then fit a line to the twist as a function of time using least absolute deviation fit to the means using the mean of the absolute deviation. We only fit the line in the time interval from $t=10$ hrs to $t=55$ hrs. The start time is the earliest time available and the final time is where the photospheric twist starts to decrease. We have measured $\alpha L/2$ to increase at about 0.021 rad/hr. 

\begin{figure}[!hc]
 \begin{center}
  \includegraphics[width=8cm]{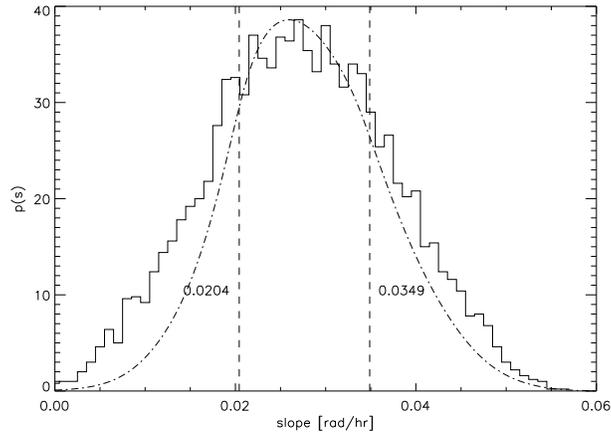}
 \end{center}
 \caption{\small{Evaluation of the uncertainty of the twist injection rate using bootstrapping method.}}
 \label{bs_fit}
 \end{figure}

We can obtain a sense of uncertainty in this measurement in several ways.  Lacking a knowledge of either the magnitude or distribution of the measurement errors we can employ the method of bootstrapping \citep{Press1986} which does not require such knowledge. We generate synthetic data sets by randomly resampling the actual data, permitting duplication, and then repeatedly fit the synthetic data using LAD on the co-temporal means.  The result is a distribution of measured rates whose central two quartiles are contained within $(0.020,\,0.035)$ rad/hr, as shown on Figure~\ref{bs_fit}. A second estimate comes from the maximum likelihood formalism.  A LAD fit is equivalent to the fit of maximum likelihood under the assumption of exponentially-distributed, additive errors with identical mean deviation equal to that of the best fit: $0.237$ rad. The central 50\% of the likelihood distribution is encompassed by an (asymmetric) range of slopes $(0.017,\,0.035)$ rad/hr, similar to that from bootstrapping. Finally, we attempted several different methods of fitting linear trends to the data including least-squares fit, fits to all data points, to the medians of each time (rather than the means), and using only A-quality points. Each method yielded a different slope, but the collection fell within the range $(0.016,\,0.038)$ rad/hr.

	
Figure~\ref{longcope_twist} shows the twist injection rate as measured by \citet{Longcope2007b} as a black curve. It also shows individual spinning (red for positive and blue for negative) and tilt (green) angles. We fit a line to the combined angle using least absolute deviations to the same time interval as on the coronal measurements and find a twist injection of about 0.0160 rad/hr (what corresponds to the -0.0160 rad/hr change in $\theta_{spin}^{+}+\theta_{spin}^{-}-2\theta_{braiding}^{+-}$, see Equation~\ref{eqn_hel_total}). 

We conclude, that the rate of change of the coronal twist is consistent with the twist injection through the photosphere. 

\section{Discussion}
\label{sec_discuss}
In the current work we have observed how helicity flows into the corona through photospheric motions. In the beginning of the time sequence coronal loops of AR 9004 appear to have negative helicity and after about 60 hours all the coronal loops appear to have positive helicity. We have observed that when a negatively twisted field is subject to the injection of helicity of positive sign, the average twist increases and eventually passes through zero (like at $t\in[40, 50]$ hrs). But this does not mean the magnetic field becomes potential. Rather, it passes through a complex non-linear equilibrium that has parts with distinct positive and parts with distinct negative twist. 

There were two weak flares associated with AR 9004 in the examined time interval, according to Solar Monitor \citep{SolarMonitorRef}. There was one C2.7 flare at $t=52.6$ hrs and then a C1.9 at $t=88.6$ hrs. \citet{Kusano2003} have proposed a model in which the field of an active region, featuring twist of both signs, undergoes magnetic reconnection. This process is accompanied by a flare and results in an untwisted magnetic field (that is, with zero helicity). The flares in AR 9004 do not seem to be associated with reliably measured drops in twist. Moreover, the AR that we study does have twist of both signs at $t\in\left[20, 55\right]$ hrs and yet does not seem to relax to a potential state at later times, after $t=72$ hrs. 

It is worth noting, however, that in the period of $t\in\left[58.9, 70,3\right]$ hrs there were almost no loops that were successfully fit. This opens the question of what might have happened at that time. It could be argued that the field indeed had relaxed to the potential state and its further positive (right-handed) twist was injected through the photosphere, but at least the apparent injection of negative helicity within about $t\in\left[55, 70\right]$ hrs (evident on Figure~\ref{longcope_twist}) suggests that this might not be the case. 

We have studied the time rate of change of the following quantity: $\langle\alpha_i L_i/2\rangle$ averaged over many reconstructed field lines. In Section~\ref{sec_method} we argued that it might be proportional to the additive self helicity of a non-linear force-free field in the similar manner as $\alpha L/2$ is proportional to twist helicity of a thin flux tube and as $\alpha \langle L_i\rangle/2$ is related to additive self helicity for a linear force-free field. 

We have found that the time rate of change of $\langle\alpha_i L_i/2\rangle$ (and arguably a generalized twist, $Tw_{gen}=H_A/\Phi^2$) is found to be about $0.021$ rad/hr. This rate is similar to the time rate of change of the flux-normalized total helicity $H_{total}/\Phi^2$ that was found to be $0.016$ rad/hr. 

The difference between $H_A$ and $H_{total}$ might be responsible for the fact that even though $H_{total}$ starts to decrease after about $t=55$hrs, the twist derived from coronal loops remains of a positive sign after $t=70$ hrs (or arguably $t=55$ hrs, as explained below) and does not seem to obviously decrease. We have noticed that a lot of bright coronal loops connecting ARs 9004 and 9002 appear at about $t=75$hrs and later at about $t=100$hrs. It is possible that while photospheric helicity injection changes sign, magnetic reconnection that happens afterward changes the balance between ``twist'' and ``writhe'' helicities, in the sense of $H_A$ as a ``generalized twist helicity'' and $H_{total}-H_A$ as a ``generalized writhe helicity''. In the current paradigm magnetic reconnection results in a decrease of magnetic energy. It is also true that linear force-free fields with larger \als are typically thought of as having larger magnetic energy than those with smaller \als \citep{Aly1992}. However the field of AR 9004 was found to be non-linear, which means, for the same system there exists a linear force-free field with lower energy \citep{Woltjer1958}. A possible scenario for the observed phenomena is the following: magnetic reconnection between ARs 9004 and 9002 lowers the total magnetic energy and possibly turns AR 9004 to a linear (or nearly linear) force-free field after about $t=70$ hrs. Within the time interval approximately $t\in[55, 70]$ hrs the total helicity of AR 9004 decreases due to photospheric motions (see Figure~\ref{longcope_twist}), however, the general trend of the self-helicity within $t\in[55, 100]$ hrs does not seem to demonstrate a similar decrease, except for maybe $t\in[60, 70]$ hrs, but the data in that range are poorly sampled; each greyscale bin has only a few data points and the means are computed for individual time frames that have even fewer points. The means seems to decrease, however, the spread in values is similar to that in $t\in[50, 60]$ hrs. Our own belief is that these two points are less reliable than the rest of the data. This is consistent with a possible interpretation of the data provided earlier in the text. This decrease in the total helicity possibly signifies a decrease a ``generalized writhe helicity'' component. The latter appears to be only a function of the shape of the domain containing field of AR 9004. So both magnetic reconnection (that has decreased the volume of this domain by reconnecting some of the magnetic flux to AR 9002) and photospheric motions might have contributed to that. 

This work was supported by NASA under grant NNX07AI01G and NSF under award ATM-0552958. We are grateful to Carolus Schrijver for the name ``(non)-linear force-free fit'' and the anonymous referee for the useful suggestion that led do a significant improve of the paper.

\clearpage
\section*{Appendix: Tangent Plane Projection}
\label{sec_appendix}
A tangent plane projection is an orthographic projection onto a plane, tangent to the Sun at the point which is called the center of the projection and the projection of the solar North directly upwards. It could easily be obtained from the plane-of-sky using the following transformations. In the starting plane-of-sky the coordinate system is assumed to be heliographic-cartesian \citep{ProjectionsThompson2006}. That is, Cartesian, with the origin at the Sun's center, $x$ axis directed towards solar West in the plane of sky, $y$ axis directed towards solar North in the plane of sky and $z$ axis directed towards the observer. The image plane is then rotated as described by Equation~\ref{proj_eqn}: first, by $-b_0$ (where $b_0$ is solar B-angle) about $x$-axis, then by $-\phi_c$ (where $\phi_c$ is the longitude of the desired projection center) about the new $y$-axis, then by $-\theta_c$ (where $\theta_c$ is the latitude of the desired projection center) about the new $x$-axis. After this sequence of rotations solar North would lie on the new $y$-axis and solar West would lie on the new $x$-axis. The last step is to perform the orthographic projection in the new $z$ direction, that is, simply to set $z$ coordinate to 0 for all points in the visible hemisphere. 

$$A_x(\theta)=\begin{pmatrix}
1 & 0 & 0 \\
0 & \cos(\theta) & -\sin(\theta) \\
0 & \sin(\theta) & \cos(\theta) 
\end{pmatrix}$$

$$A_y(\phi)=\begin{pmatrix}
\cos(\phi) & 0 & \sin(\phi) \\
0 & 1 & 0 \\
-\sin(\phi) & 0 & \cos(\phi) 
\end{pmatrix}$$

\begin{equation}A=A_x(b_0)A_x(-\theta_c)A_y(-\phi_c)A_x(-b_0)\label{proj_eqn}\end{equation}

\begin{figure}[!hc]
 \begin{center}
  \includegraphics[width=4.cm]{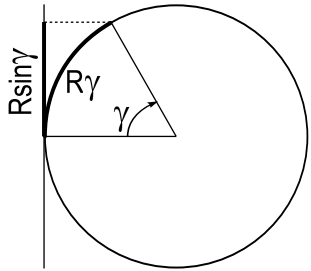}
 \end{center}
 \caption{\small{Distortions of distance on tangent plane projection are small if the size of the extracted region is small. For example, if a distance between the point of tangency and another point on the Sun is $R\gamma$, then (if $\gamma\le\pi/2$) the distance between the center of the projection and the projection of that point is $R\sin\gamma$.}}
 \label{proj_discrep}
 \end{figure}

This projection is non-conformal (it does distort local shape) and distorts the distances. But those distortions are independent of the location of the point of tangency and only increase with the size of the desired box. For example (see Figure~\ref{proj_discrep}), if one considers a point on the sphere, which radius-vector from Sun's center $\rvec$ makes an angle $\gamma$ with the radius-vector from Sun's center to the projection center, $\rvec_0$, then the distance on the sphere between $\rvec$  and $\rvec_0$ is $R\gamma$ and the distance between the projections of these two points on the tangent plane is $R\sin\gamma$. For $\gamma=30^\circ$ (or the box of $60^\circ$, which is enough to fit most active regions), the foreshortening factor would be about 5\%. 

For purposes of tracking long time sequences and dealing with off-center solar images, heliographic coordinates (with longitude $\phi$ along horizontal axis and latitude $\theta$ along vertical) are often used, with the longitude either equally spaced along the vertical axis \citep{ProjectionsThompson2006} or with spacing that changes with the distance from the equator \citep{Welsch2009}. The first one is commonly referred in cartography as plate Carr\'ee projection and a particular case of the second one frequently used in Solar Physics is called Mercator projection \citep{projection_textbook}. 

\begin{figure}[!hc]
 \begin{center}
  \begin{tabular}{p{5.5cm}p{5.5cm}p{5.5cm}}
   \includegraphics[width=8.cm]{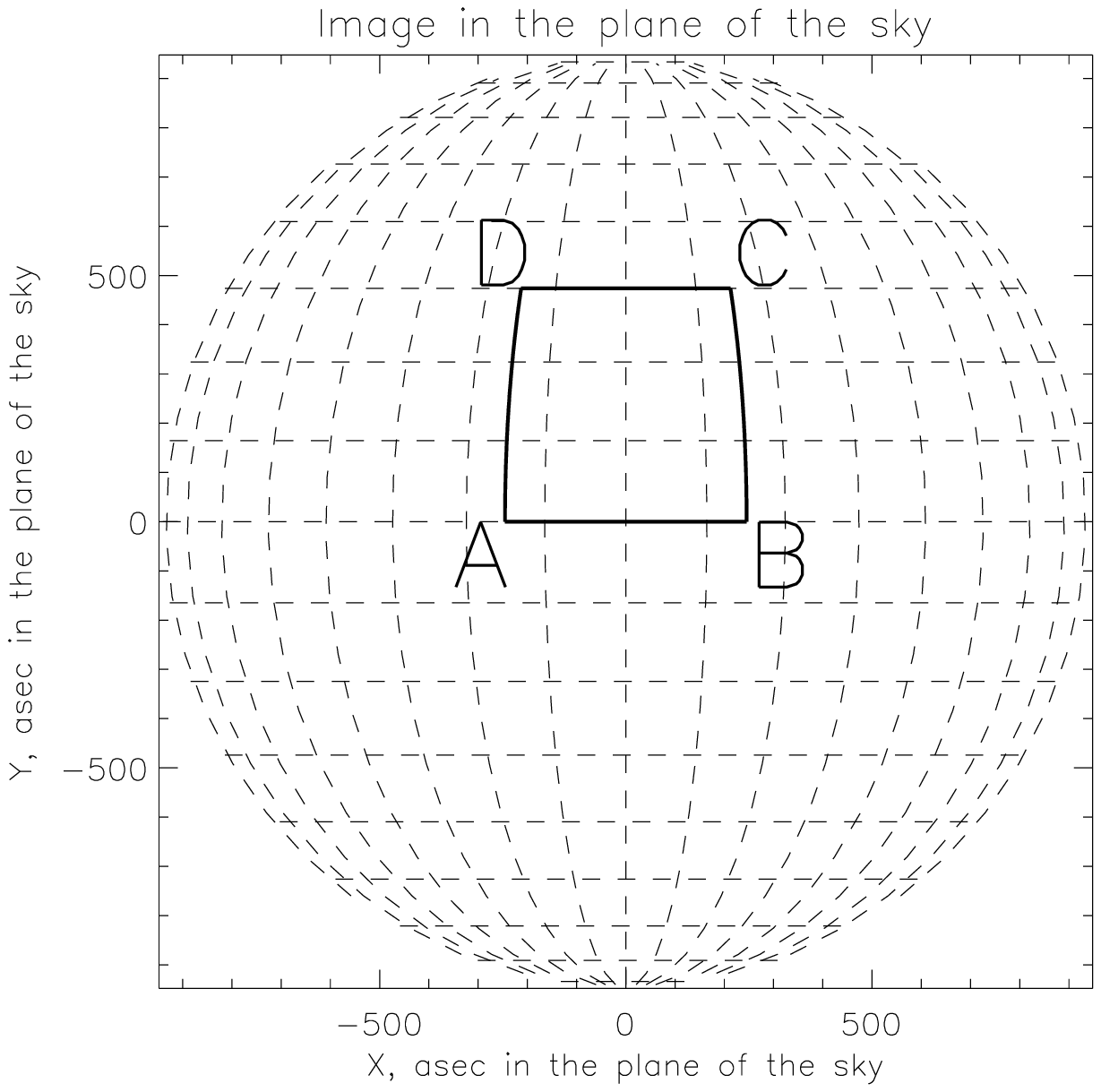} &
   \includegraphics[width=8.cm]{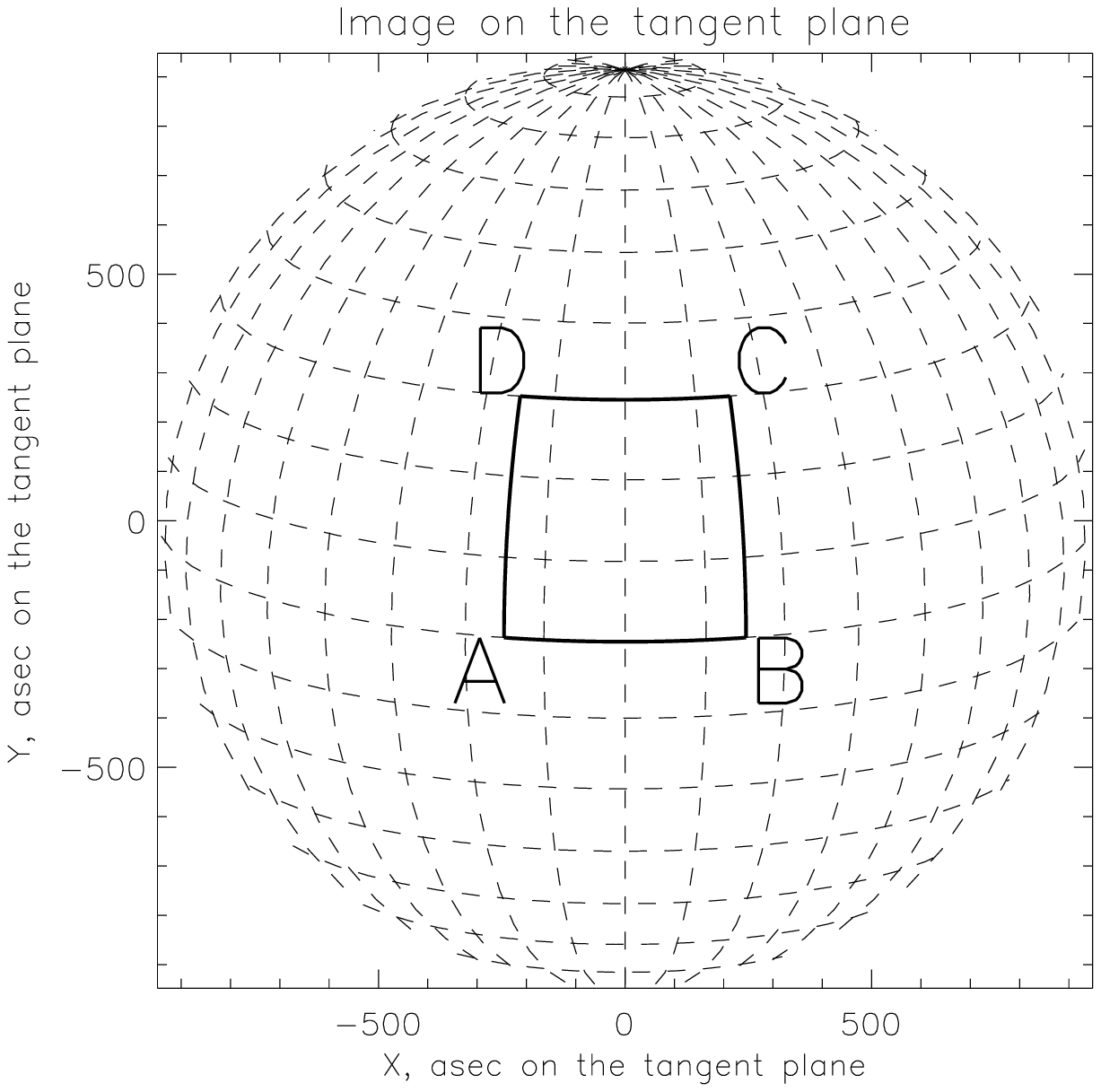} &
   \includegraphics[width=8.cm]{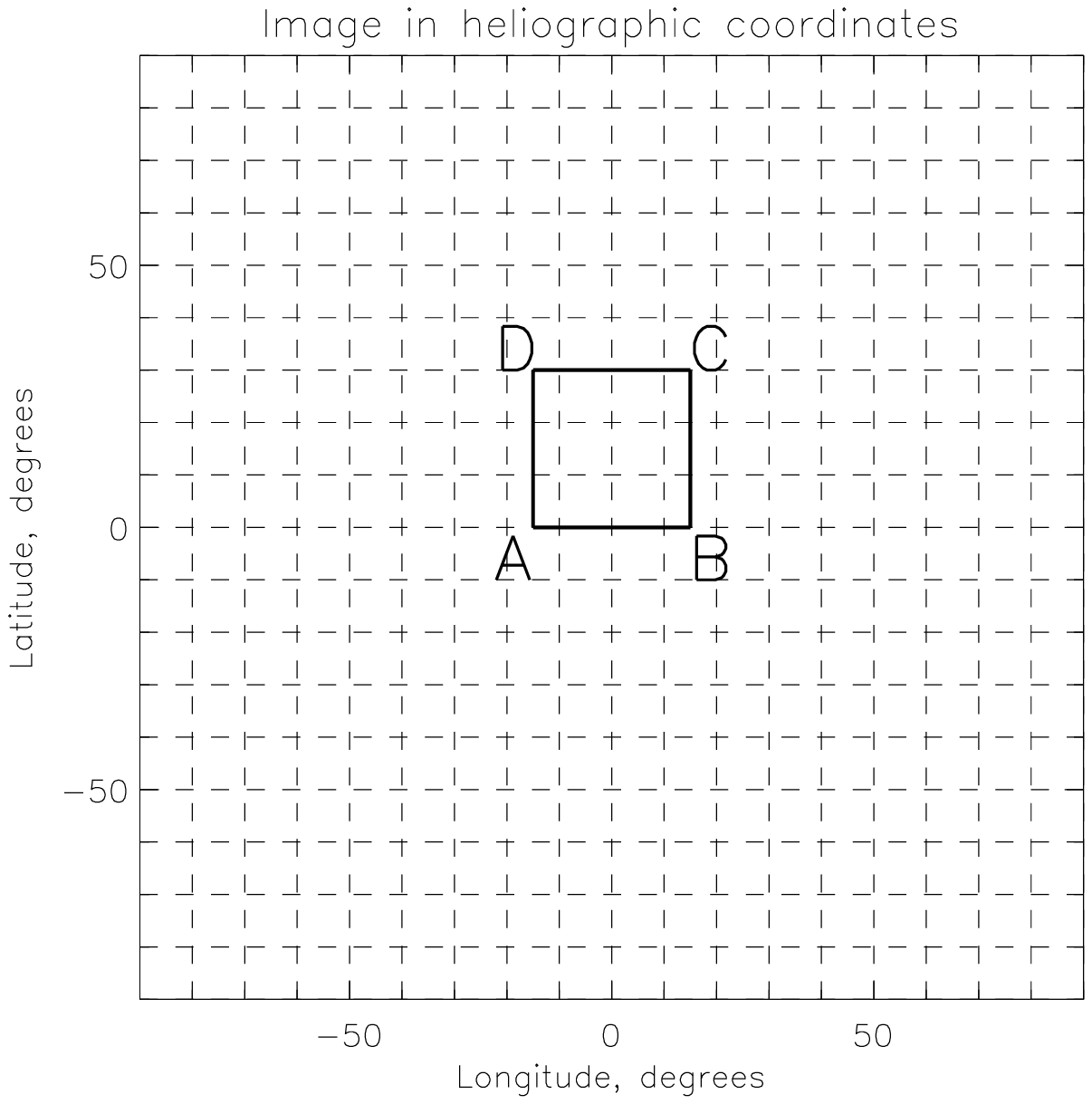} \\
  \end{tabular}
 \end{center}
 \caption{\small{Illustration of heliocentric-cartesian (orthographic) and heliographic projections. $ABCD$ is a $30^{\circ}\times 30^{\circ}$ ``square'', centered at $0^{\circ}$W, $15^{\circ}$N (roughly the size and position of ARs 9002 and 9004 when they pass through the central meridian) in three different projections. \textit{(Left)} -- in the plane of the sky (orthographic projection, centered at $0^{\circ}$W, $0^{\circ}$N and neglecting the $b$-angle for illustrative purposes), \textit{(middle)} -- in the tangent plane (orthographic projection, centered in the middle of $ABCD$, i.e., $0^{\circ}$W, $15^{\circ}$N), \textit{(right)} -- in heliographic coordinates (plate Carr\'ee projection). Dashed lines are lines of constant latitude and longitude.}}
 \label{projections}
 \end{figure}

A tangent plane projection, such as we use, is somewhat less common. The first reason why we chose it is that the distortions are independent of latitude (unlike for Mercator or plate Carr\'ee projections, which have systematic latitude-dependent errors), as explained above. The second reason why we choose the tangent plane projection is that it is distorts global shapes less than cylindrical projections. Since the shape of a coronal loop is crucial for determining its \al, we believe it is more important to preserve shapes globally rather than locally, and orthographic projections are better in this sense than cylindrical ones. For example, let us consider a $30^{\circ}\times 30^{\circ}$ ``square'' $ABCD$, centered at $0^{\circ}$W, $15^{\circ}$N (roughly the size and position of ARs 9002 and 9004 when they pass through the central meridian), as shown on Figure~\ref{projections}. The lengths of the arches on the sphere are $AB=BC=AD=\frac{\pi}{6}R_\sun\approx0.524R_\sun$ and $CD=\frac{\pi}{6}\cos\frac{\pi}{6}R_\sun\approx0.454R_\sun$, so $CD/AB\approx0.86$. In a cylindrical projection $CD/AB=1$ and on a tangent plane centered at $0^{\circ}$W, $15^{\circ}$N, $CD/AB\approx0.87$ (the latter is obtained by writing parametric equations for the projections of the box's sides and computing their length using standard methods). The foreshortening factors on the tangent plane are about the same for all four sides and are about $0.99$. We thus believe that by using tangent plane projection in this particular case we make about 1\% error due to the projection effect, as opposed to about 15\% error in a cylindrical projection (given for $CD/AB$; the ratio $AB/BC$ would differ depending on the particular type of the cylindrical projection).

\bibliography{c:/localtexmf/bib/short_abbrevs,c:/localtexmf/bib/full_lib,c:/localtexmf/bib/my_bib}
\end{document}